\documentclass[longauth]{aa} % for the long lists of affiliations 

\usepackage{url}
\usepackage{graphicx}
\usepackage[varg]{txfonts}
\usepackage{hyperref} %% breaklinks=true to avoid \citeads line fills
\hypersetup{breaklinks=true,colorlinks=true,linkcolor=blue,citecolor=blue,urlcolor=blue}
\usepackage{natbib,twoopt}
\bibpunct{(}{)}{;}{a}{}{,}
\usepackage{color}
\usepackage{deluxetable}
\usepackage{pdflscape}

\graphicspath{{figures/}}

\newcommand{\kms}{km\,s$^{-1}$}

\newcommand{\kmsM}{km\,s$^{-1}$\,Mpc$^{-1}$}
\newcommand{\Msun}{M$_\odot$}

\newcommand{\califa}{\texttt{CALIFA}} % CALIFA survey
\newcommand{\sauron}{\texttt{SAURON}}      % SAURON kinematics
\newcommand{\atlas}{\texttt{ATLAS$^\mathrm{3D}$}} % ATLAS3D project
\newcommand{\eps}{\ensuremath{\varepsilon}}
\newcommand{\epse}{\ensuremath{\varepsilon_e}}

\newcommand{\lamR}{\ensuremath{\lambda_\mathrm{R}}}
\newcommand{\lamRe}{\ensuremath{\lambda_\mathrm{Re}}}
\newcommand{\lamtwoRe}{\ensuremath{\lambda_\mathrm{2\,Re}}}
\newcommand{\lamhalfRe}{\ensuremath{\lambda_\mathrm{0.5\,Re}}}
\newcommand{\VSe}{\ensuremath{(V/\sigma)_\mathrm{e}}}

\newcommand{\Mstar}{\ensuremath{M_\star}}

\newcommand{\epsintr}{\ensuremath{\varepsilon_{90^\circ}}}

\newcommand{\VSintr}{\ensuremath{(V/\sigma)_{90^\circ}}}
\newcommand{\lamRintr}{\ensuremath{\lambda_{\mathrm{R},{90^\circ}}}}
\newcommand{\lamReintr}{\ensuremath{\lambda_{\mathrm{Re},{90^\circ}}}}

\newcommand{\iav}{\ensuremath{i_\mathrm{av}}}

\newcommand{\Reff}{R$_\mathrm{e}$}

\def\todo[#1]#2{\noindent {\color{blue} {\bf[#1]:} #2}}

\begin{document} 

%=====================================================================
% FRONT MATTER
%=====================================================================

\titlerunning{Stellar angular momentum in the CALIFA survey}
\authorrunning{Falc{\'o}n-Barroso et al.}

\title{The CALIFA view on stellar angular momentum across the Hubble sequence}

\author{J.~Falc\'on-Barroso\inst{\ref{inst1},\ref{inst2}}\thanks{Email: jfalcon@iac.es}
   \and G.~van~de~Ven\inst{\ref{inst3}}
   \and M.~Lyubenova\inst{\ref{inst4}}
   \and J.~Mendez-Abreu\inst{\ref{inst1},\ref{inst2}}
   \and J.~A.~L.~Aguerri\inst{\ref{inst1},\ref{inst2}}
   \and B.~Garc\'ia-Lorenzo\inst{\ref{inst1},\ref{inst2}}
   \and S.~Bekerait{\'e}\inst{\ref{inst5}}
   \and S.~F.~S\'anchez\inst{\ref{inst6}}
   \and B.~Husemann\inst{\ref{inst7}}
   \and R.~Garc{\'\i}a-Benito\inst{\ref{inst8}}
   \and R.~M.~Gonz\'alez Delgado \inst{\ref{inst8}}
   \and D.~Mast\inst{\ref{inst9},\ref{inst10}}
   \and C.J.~Walcher\inst{\ref{inst5}}
   \and S.~Zibetti\inst{\ref{inst11}}
   \and L. Zhu\inst{\ref{inst12}}
%-----------------------------------------------------
   \and J.~K.~Barrera-Ballesteros\inst{\ref{inst6}}
   \and L.~Galbany\inst{\ref{inst13}}
   \and P.~S\'anchez-Bl\'azquez\inst{\ref{inst14}}
   \and R.~Singh\inst{\ref{inst7}}
   \and R.~C.~E.~van den Bosch\inst{\ref{inst7}}
   \and V.~Wild\inst{\ref{inst15}}
%-----------------------------------------------------
   \and J.~Bland-Hawthorn\inst{\ref{inst16},\ref{inst17}}
   \and R.~Cid~Fernandes\inst{\ref{inst18}}
   \and A.~de~Lorenzo-C{\'a}ceres\inst{\ref{inst1},\ref{inst2}}
   \and A.~Gallazzi\inst{\ref{inst11}}
   \and R.~A.~Marino\inst{\ref{inst19}}
   \and I.~M\'arquez\inst{\ref{inst8}}
   \and R.~F.~Peletier\inst{\ref{inst20}}
   \and E.~P\'erez\inst{\ref{inst8}}
   \and I.~P\'erez\inst{\ref{inst21},\ref{inst22}}
   \and M.~M.~Roth\inst{\ref{inst5}}
   \and F.~F.~Rosales-Ortega\inst{\ref{inst23}}
   \and T.~Ruiz-Lara\inst{\ref{inst1},\ref{inst2}}
   \and L.~Wisotzki\inst{\ref{inst5}}
   \and B.~Ziegler\inst{\ref{inst3}}
} 
\institute{
Instituto de Astrof\'isica de Canarias, V\'ia L\'actea s/n, E-38205 La Laguna, Tenerife, Spain\label{inst1}
\and
Departamento de Astrof\'isica, Universidad de La Laguna, E-38205 La Laguna, Tenerife, Spain\label{inst2}
\and
Department of Astrophysics, University of Vienna, T\"urkenschanzstrasse 17, 1180 Vienna, Austria\label{inst3}
\and 
European Southern Observatory, Karl-Schwarzschild-Str. 2, 85748 Garching b. M\"unchen, Germany\label{inst4}
\and
Leibniz-Institut f\"ur Astrophysik Potsdam (AIP), An der Sternwarte 16, D-14482 Potsdam, Germany,\label{inst5}
\and
Instituto de Astronom\'ia, Universidad Nacional Aut\'onoma de M\'exico, Apartado Postal 70-264, M\'exico D.F., 04510 M\'exico\label{inst6}
\and
Max-Planck-Institut f\"ur Astronomie, K\"onigstuhl 17, D-69117 Heidelberg, Germany\label{inst7}
\and 
Instituto de Astrof\'isica de Andaluc\'ia (IAA/CSIC), Glorieta de la Astronom\'ia s/n Aptdo. 3004, E-18080 Granada, Spain\label{inst8}
\and 
Observatorio Astron\'omico, Laprida 854, X5000BGR, C\'ordoba, Argentina.\label{inst9}
\and 
Consejo de Investigaciones Cient\'{i}ficas y T\'ecnicas de la Rep\'ublica Argentina, Avda. Rivadavia 1917, C1033AAJ, CABA, Argentina.\label{inst10}
\and 
INAF-Osservatorio Astrofisico di Arcetri - Largo Enrico Fermi, 5 - I-50125 Firenze, Italy\label{inst11}
\and
Shanghai Astronomical Observatory, Chinese Academy of Sciences, 80 Nandan Road, Shanghai 200030, China\label{inst12}
\and
PITT PACC, Department of Physics and Astronomy, University of Pitts-burgh, Pittsburgh, PA 15260, USA\label{inst13}
\and
Departamento de F\'{\i}sica Te\'orica, Universidad Aut\'onoma de Madrid, E-28049, Madrid, Spain\label{inst14}
\and
School of Physics and Astronomy, University of St Andrews, North Haugh, St Andrews, KY16 9SS, UK (SUPA)\label{inst15}
\and
Sydney Institute for Astronomy, School of Physics A28, University of Sydney, NSW 2006, Australia\label{inst16}
\and
ARC Centre of Excellence for All-sky Astrophysics in 3D (ASTRO-3D)\label{inst17}
\and
Departamento de F\'{\i}sica, Universidade Federal de Santa Catarina, P.O. Box 476, 88040-900, Florian\'opolis, SC, Brazil\label{inst18}
\and
ETH Z\"urich, Institute for Astronomy, Wolfgang-Pauli-Str. 27, 8093 Z\"urich, Switzerland\label{inst19}
\and
Kapteyn Astronomical Institute, University of Groningen, Postbus 800, NL-9700 AV Groningen, the Netherlands\label{inst20}
\and
Departamento de F\'{\i}sica Te\'orica y del Cosmos, University of Granada, Facultad de Ciencias (Edificio Mecenas), E-18071 Granada, Spain\label{inst21}
\and
Instituto Carlos I de F\'{\i}sica Te\'orica y Computaci\'on\label{inst22}
\and
Instituto Nacional de Astrof{\'i}sica, {\'O}ptica y Electr{\'o}nica, Luis E. Erro 1, 72840 Tonantzintla, Puebla, Mexico\label{inst23}
}

%\date{~\hfill\fbox{\textbf{\emph{DO NOT DISTRIBUTE}}}}
\date{Received July 29, 2019; accepted October 11, 2019}

\abstract{We present the apparent stellar angular momentum over the optical 
extent of 300 galaxies across the Hubble sequence, using integral-field 
spectroscopic (IFS) data from the CALIFA survey. Adopting the same \lamR\ parameter 
previously used to distinguish between slow and fast rotating early-type 
(elliptical and lenticular) galaxies, we show that spiral galaxies as expected 
are almost all fast rotators. 
Given the extent of our data, we provide relations for \lamR\ measured in different 
apertures (e.g. fractions of the effective radius: 0.5\,\Reff, \Reff, 2\,\Reff), including 
conversions to long-slit 1D apertures.
Our sample displays a wide range of \lamRe\ values, consistent with previous IFS studies.
The fastest rotators are dominated by relatively massive and highly star-forming Sb 
galaxies, which preferentially reside in the main star-forming sequence. These galaxies reach 
\lamRe\ values of $\sim$0.85, are the largest galaxies at a given mass, and display some of the
strongest stellar population gradients. Compared to the population of S0 galaxies, our findings 
suggest that fading may not be the dominant mechanism transforming spirals into lenticulars.
Interestingly, we find that \lamRe\ decreases for late-type Sc and Sd spiral galaxies, with 
values than in occasions puts them in the slow-rotator regime. While for some of them this 
can be explained by their irregular morphologies and/or face-on configurations, others are 
edge-on systems with no signs of significant dust obscuration. The latter are typically at 
the low-mass end, but this does not explain their location in the classical $(V/\sigma,\eps)$ 
and ($\lamRe,\eps$) diagrams. Our initial investigations, based on dynamical models, suggest 
that these are dynamically hot disks, probably influenced by the observed important fraction 
of dark matter within \Reff.}

\keywords{Galaxies: kinematics and dynamics -- 
          Galaxies: elliptical and lenticular, cD -- 
          Galaxies: spiral -- 
          Galaxies: structure -- 
          Galaxies: evolution -- 
          Galaxies: formation}
         
\maketitle

%=====================================================================
% BEGIN PAPER
%=====================================================================

%=====================================================================
\section{Introduction}
\label{S:intro}
%=====================================================================

After its mass, one of the key parameters that determine the fate of a galaxy is 
its angular momentum.  A robust result from cosmological simulations is that the 
angular momentum distribution of dark matter halos is nearly constant with 
redshift \citep[e.g.][]{bullock01}. The amount of angular momentum that is being 
transferred to the baryons is then believed to set the size of galactic disks 
\citep{mo98} and to form the basis for the mass-size relation of galaxies 
\citep{shen03}. At the same time, tidal interactions and in particular mergers 
between galaxies can disturb or even fully destroy the disk so that the memory 
of the initial angular momentum might well be lost \citep[e.g.][]{toomre72}.

Galaxy mergers are indeed believed to be an important reason why 
spheroid-dominated galaxies with surface brightness profiles close to 
de Vaucouleur (with a S\'ersic index $n$\,$\sim$\,4)  deviate from the mass-size 
relation of galaxies with outer surface brightness profiles close to exponential 
($n$\,$\sim$\,1). The latter include disk-dominated spiral 
galaxies, but the relation seems to extend toward lower masses, including dwarf 
elliptical galaxies \citep[e.g.][]{kormendy12} and possibly even down to the dwarf 
spheroidal galaxies \citep[e.g.][]{brasseur2011}.\looseness-2

Even though stellar rotation is observed in dwarf elliptical galaxies 
\citep[e.g.][]{toloba11} and possibly even in dwarf spheroidal galaxies 
\citep[e.g.][]{battaglia08}, the motion of their stars remains dominated by 
dispersion. This implies that the initial angular momentum that set their sizes 
got reduced, likely as a result of the mechanisms that are believed to have 
transformed dwarf disk galaxies into these dwarf spheroid galaxies. Transformation 
mechanisms proposed such as tidal interaction and ram pressure 
stripping are thought to act stochastically, as reflected in the large diversity 
in photometric, kinematic and stellar population properties 
\citep[e.g.][]{rys1,rys2,rys3}, but sudden dramatic changes as a result of, for 
example, mergers are expected to be rare \citep{amorisco14}. 

This shows that even if the process of transferring the angular momentum from 
halo to disk is broadly understood \citep[e.g.][]{burkert16}, there might not be anymore a 
direct link between the size of the disk of galaxy and its current stellar 
angular momentum. However, the comparison between current angular momentum and 
size of galaxies at a given mass, provides constraints on the changes in angular 
momentum and on the mechanisms that caused these changes. The latter 
mechanisms, in turn, are in all likelihood the same that are responsible for 
defining the Hubble sequence of galaxies \citep[e.g.][]{af12}. Clearly, a crucial ingredient in 
uncovering the evolution of galaxies is a homogeneous and statistically-sound 
census of the stellar angular momentum in nearby galaxies of all Hubble types. 

The \sauron\ project \citep{dezeeuw02} combined the observed stellar 
line-of-sight velocity and dispersion fields of 48 early-type galaxies to 
compute the parameter $\lamRe$ as a measure of the apparent stellar angular 
momentum within one effective radius \Reff\ \citep{emsellem07}. The \atlas\ 
survey \citep{cappellari_etal_2011} extended this to a volume-limited sample of 
260 early-type galaxies out to 42\,Mpc to confirm the existence of two families: 
slow rotators elliptical galaxies with complex stellar velocity fields and fast 
rotator lenticular as well as elliptical galaxies with regular stellar velocity 
fields \citep{krajnovic_etal_2011}. With the advent of new 2D surveys (e.g. 
SAMI, \citealt{sami}; SLUGGS, \citealt{arnold14}; MASSIVE, \citealt{ma14}; 
MaNGA, \citealt{manga}), there has been steady progress in this 
field over the past decade. While initial samples were still biased towards 
early-type systems \citep[e.g.][]{arnold14, fogarty14, veale17a}, the topic has 
remained active and has spurred the study of angular momentum in 
even larger samples of galaxies (including spirals) by the SAMI (\citealt{cortese16}, 
hereafter C16; \citealt{vdsande17b}, hereafter vdS17) and MaNGA (\citealt{graham18}, 
hereafter G18) survey teams.\looseness-2

The CALIFA survey \citep{Sanchez_etal_2012} of a diameter-selected 
sample of up to 600 nearby galaxies is providing stellar velocity and dispersion 
fields that not only extend further out in radius, but also covers galaxies of 
all Hubble types. The goal of this paper is to use the stellar velocity and dispersion 
maps of 300 observed CALIFA galaxies presented in \citet[][hereafter FLV17]{flv17} to 
provide a robust census of the apparent stellar angular momentum across the 
Hubble sequence and investigate the properties of the galaxies in some of the most 
extreme regions of the ($\lamRe,\eps$) diagram. Part of the results 
shown here were already presented in \citet{fb15}, and have been used in the
recent literature for comparison with other surveys \citep[e.g.][]{cap16,schulze18} or 
for highlighting the peculiarities of certain types of galaxies \citep[e.g.][]{delMoral19}.

The paper is organised as follows. After describing in section~\ref{sec:data} the available 
data for 300 galaxies, we present in \S\ref{sec:lamR} the resulting apparent stellar angular 
momentum, within apertures of different radii and when only long-slit data would 
be available, as radial profiles. We place in section~\ref{sec:lamRepsdiagram} 
the galaxies on the $(V/\sigma,\eps)$ and ($\lamRe,\eps$) diagrams to 
investigate the rotational versus pressure support homogeneously among galaxies 
of all morphologies, showing trends among types and discussing their relation to other global
parameters. We conclude in \S\ref{sec:conclusions}. Appendix~\ref{app:deproj} describes the 
procedures used to deproject our \lamRe\ measurements. Table~\ref{tab:sample} 
provides all the quantities used and derived in our study. Throughout we adopt 
$H_0=70$\,\kmsM, $\Omega_M=0.3$ and $\Omega_L=0.7$ for respectively the Hubble 
constant, the matter density and the cosmological constant, although these 
parameters only have a small effect on the physical scales of the galaxies due 
to their relative proximity.\looseness-2

%=====================================================================
\section{CALIFA IFU and ancillary data} 
\label{sec:data}
%=====================================================================

The Calar Alto Legacy Integral Field Area (\califa) survey was the first 
integral-field spectroscopic (IFS) survey of a diameter selected ($45\arcsec < 
D_{25} < 80\arcsec$) sample of up to 600 galaxies in the local universe ($0.005 
< z < 0.03$) of all Hubble types \citep{Sanchez_etal_2012}. The so-called CALIFA 
`mother sample' of 938 galaxies, from which targets are randomly observed based 
on visibility, is representative in stellar mass over two orders of magnitude 
$9.4 < \log(M_\star/M_\odot) < 11.4$. This means that after a straightforward 
volume correction based on $D_{25}$, the mass (and corresponding luminosity) 
function over this range is recovered to better than 95\%  \citep{Walcher_etal_2014}. 
The $65\arcsec\times72\arcsec$ field-of-view of the employed PMAS/PPAK-IFU 
\citep{Kelz_etal_2006} covers the full optical extent of the selected galaxies, with 
a complete filling factor achieved through a three-point dithering scheme, and 
with a spatial sampling of 1\arcsec\ that over-samples the spatial resolution by 
about a factor three \citep{Husemann_etal_2013}. The typical Pont-Spread-Function 
size is FWHM$\sim$2.5\arcsec\ \citep{sanchez16}, that corresponds to an average 
physical resolution of 0.7\,kpc and a range betweeen 0.2-1.5\,kpc within the considered 
redshift range.

In this paper, we use the high-quality stellar kinematics presented in FLV17 
from the V1200 dataset. Briefly,  stellar velocity ($V$) and velocity dispersion ($\sigma$) 
maps were computed using the pPXF code of \citet{ppxf}, after the data had been 
Voronoi binned \citep{Cappellari_Copin_2003} to a signal-to-noise ratio (SNR) of 20 per pixel. 
We used as templates the Indo-US spectral library \citep{indous} over the wavelength 
range covered by the V1200 grating (i.e. 3750$-$4550\,\AA), that includes prominent 
absorption features such as Ca H+K, H$\gamma$ or the H$\delta$ lines. The typical relative
uncertaininties of our measurements are $\sim$5\% for $\sigma\ge150$\,\kms. Below that value
they increase up tp 50\% for velocity dispersions as low as 20\,\kms. We refer the reader to FLV17
for more details.

Additional global galaxy properties used here are: (i) distances based on redshift with 
Hubble flow corrected for Virgo infall (see \citealt{Walcher_etal_2014}); (ii) SDSS redshifts, apparent 
magnitudes and corresponding colors; (iii) light concentrations based on SDSS $r$-band 50 and 90 
percentile Petrosian radii; (iv) effective radii (\Reff) estimated using a growth-curve analysis applied to 
the SDSS images as described in \citet{Walcher_etal_2014}; (v) stellar masses based on \textit{Sunrise} 
spectral energy distribution fits from \citet{Walcher_etal_2014}; (vi) global star formation rates (SFRs) 
based on Balmer-decrement corrected H$\alpha$ fluxes extracted from the CALIFA datacubes 
\citep{sanchez17}; (vii) stellar population parameters (average ages and age gradients) 
from \citet{rgb17} using CALIFA data Voronoi binned to reach a target SNR$\sim$20. The 
resulting spectra of each bin was then processed using PyCASSO \citep{cid_fernandes13,amorin17} 
using a combination of the GRANADA  \citep{rosa05} and MILES models \citep{vazdekis15} respectively. 
Reported ages are averages within \Reff, while radial age gradients were computed performing a 
robust linear fit over the entire inner \Reff.

%...............................................................................
\begin{figure}
\begin{center}
\includegraphics[width=\linewidth]{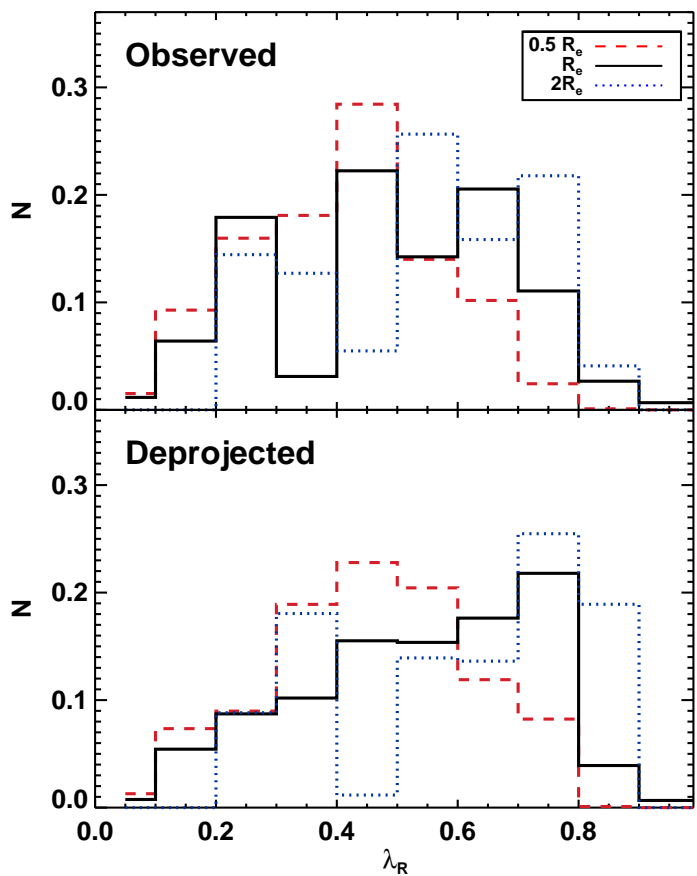}
\caption{Volume-corrected \lamR\ distributions for different aperture sizes 
(i.e. 0.5\,\Reff, \Reff, and 2\,\Reff). Top panel shows the distribution of 
\lamR\ as observed, while the bottom panel presents the distribution of 
deprojected values (as explained in Appendix~\ref{app:deproj}).}
\label{fig:lamdistrib}
\end{center}
\end{figure}
%...............................................................................

%...............................................................................
\begin{figure}
\begin{center}
\includegraphics[width=\linewidth]{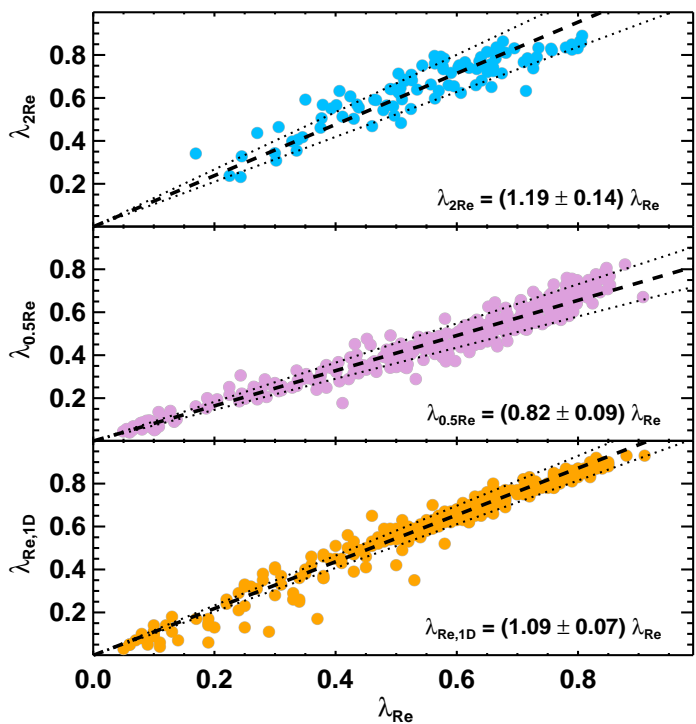}
\caption{Stellar angular momentum ($\lamR$) aperture relations for different 
aperture sizes. Top and middle panels show the relation between \lamRe\ with the 
values of smaller/larger apertures 0.5\,\Reff\ and 2\,\Reff. The bottom panel 
presents the comparison of \lamRe\ with that computed with a 1D long-slit along 
the major axis of the galaxies. The dashed lines indicate the biweight mean 
values while dotted lines the standard deviation.}
\label{fig:aper_relations}
\end{center}
\end{figure}
%...............................................................................

%=====================================================================
\section{Apparent stellar angular momentum}
\label{sec:lamR}
%=====================================================================

Following \citet{emsellem07}, we define the apparent stellar angular momentum as 
\begin{equation}
  \label{eq:deflamR}
  \lambda_R = \frac{\sum_j^N F_j R_j |V_j|}{\sum_j^N F_j R_j \left(V_j^2 + \sigma_j^2\right)^\frac12}
\end{equation}

\noindent where $F_j$, $R_j$, $V_j$ and $\sigma_j$ are the flux, polar
radius, velocity, and dispersion per spatial bin $j$ for which the centroid falls 
within an elliptic aperture with given semi-major axis $R$, ellipticity $\eps$ 
and position angle $PA$.

We adopt for $\eps$ and $PA$ the median values of the outer 10\% radial points 
of respectively the ellipticity and position angle profile resulting from an 
IRAF \textit{ellipse} model of the SDSS $r$-band image of each galaxy 
\citep{jairo17}. This is in contrast to previous studies 
\citep[e.g.][]{emsellem07}, where the mean ellipticity within \Reff\ was used 
instead (\epse). We decided on this option as the effect of non-axisymmetric 
distortions such as caused by bars, spiral arms and tidal interactions are 
minimized --- supported also by the close correspondence between the kinematic 
position angle based on the stellar velocity fields and the latter $PA$ based on 
the outer radii (see Fig.~2 of \citealt{barrera14}). We have estimated that ellipticities 
measured within \Reff\ are on average 6\%\ smaller than those used here. Nevertheless, 
we confirm  the good correspondence between the different approaches, except for 
extreme cases.

%---------------------------------------------------------------------
\subsection{Global values and aperture transformations}
\label{sec:lamRglobal}
%---------------------------------------------------------------------

Our dataset allows the exploration of the specific angular momentum on different 
aperture sizes. The vast majority of galaxies reach \Reff\ (97\%), while 61\%\ reach 
up to 2\,\Reff\ (see Fig.~4 in FLV17 for details). 

\begin{figure}
\begin{center}
\includegraphics[width=\linewidth]{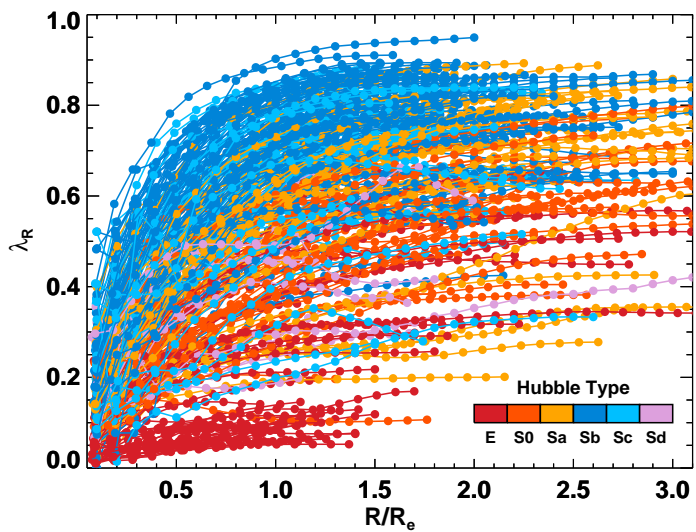}
\caption{Integrated \lamR\ profiles for our CALIFA sample of 300 galaxies. The profiles 
are normalized with \Reff\ and color-coded by Hubble type (as indicated by the colorbar).}
\label{fig:profiles}
\end{center}
\end{figure}

Figure~\ref{fig:lamdistrib} shows the normalised observed and deprojected \lamR\ 
distributions measured within 0.5\,\Reff, \Reff, and 2\,\Reff. For the 8 galaxies 
not reaching one \Reff\ with our signal-to-noise requirements, we have extrapolated 
their values up to \Reff\ based on their integrated profiles, as this extrapolation would 
be relatively safe (see Fig.~\ref{fig:profiles}). Note that we did not attempt to extrapolate 
values up to 2\,\Reff, as it would be more uncertain. Deprojected \lamR\ values were 
obtained following the prescriptions outlined in Appendix~\ref{app:deproj}. In order to 
provide the most representative distributions for the general population of galaxies, the 
histograms have been computed weighting each galaxy contribution by the volume 
correction factor (V$_{\rm max}^{-1}$). The figure shows a mild increase in \lamR\ with 
the aperture size, as expected if the majority of galaxies display clear rotation. 
While this difference may not be so obvious in the distributions of observed 
values, it shows clearly in the deprojected ones, peaking at $\sim$\,0.45, 
$\sim$\,0.75, and $\sim$\,0.80 respectively for each aperture. 

Since data reaching one \Reff\ is not always available in other data sets, we provide 
here transformations between apertures based on our data. This enables us to 
investigate how well the measured apparent stellar angular momentum at smaller 
radii can be extrapolated to larger radii. In Fig.~\ref{fig:aper_relations} we 
compare \lamRe\ with \lamhalfRe\ and \lamtwoRe\ measured within half and twice 
the half-light radius respectively, for those galaxies for which the kinematics 
extends far enough. The dashed curves represent the biweight mean relations:
\begin{equation}
  \label{eq:lamRetwice}
  \lamhalfRe = (0.82 \pm 0.09) \; \lamRe, 
  \qquad
  \lamtwoRe = (1.19 \pm 0.14) \; \lamRe,
\end{equation}
which provide approximate extrapolations for galaxies of all Hubble types. Note 
that the systematic trend discussed in Fig.~\ref{fig:lamdistrib} is even more 
evident here. Also, the lack of low \lamRe\ and \lamtwoRe\ values in the top panel 
highlights one of the limitations of the CALIFA target selection: large and 
massive nearby early-type galaxies, which are the main constituents of the slow 
rotator family \citep[e.g.][]{emsellem_etal_2011, veale17b} appear in low numbers. 
Nevertheless, the correlations presented here are in good agreement with those 
presented in \citet{vdsande17b} (e.g. \lamhalfRe$\approx$\,0.79\,\lamRe).

The availability of stellar kinematic maps is rapidly increasing with ongoing 
and upcoming integral-field spectroscopic instruments and surveys. Even so, much 
of the stellar kinematic data at higher redshift will remain based on long-slit 
spectroscopy, which instead provides stellar velocity and dispersion profiles. 
Assuming the usual major-axis orientation of the long-slit, we use the 
\emph{kinemetry} routine \citep{davor06} to extract from the stellar kinematic 
maps of all CALIFA galaxies a major-axis rotation and dispersion profile. In the 
same way as equation~\eqref{eq:deflamR} for $\lamRe$, we then compute 
$\lambda_\mathrm{Re,1D}$ from all radial bins out to the half-light radius 
$R_e$, resulting in the correlation shown in the bottom panel of 
Fig.~\ref{fig:aper_relations}. The solid curve represents the biweight mean 
relation
\begin{equation}
  \label{eq:lamReslit}
  \lambda_\mathrm{Re,1D} = (1.09 \pm 0.07) \; \lamRe.
\end{equation}
Our relation differs somewhat from \citet{toloba15} 
(i.e. $\lambda_\mathrm{Re,1D}\approx1.56$\,\lamRe), likely 
due to differences in the size and type of galaxy samples used: 300 galaxies
of all Hubble types versus 39 dwarf elliptical galaxies in the Virgo cluster.
In addition, to aid the comparison with high redshift measurements, we 
computed the relation between \lamRe\ and (V/$\sigma$)$_\mathrm{Re,1D}$. As shown
in Eq.~B1 of \citet{emsellem_etal_2011}, the relation between \lamRe\ and (V/$\sigma$) has a 
quadratic form depending on a single parameter $\kappa$. We have fitted the relation 
and obtained a value of $\kappa=1.1$ for all Hubble types, which is the same value derived 
in the ATLAS3D survey for early-type galaxies.\looseness-1

%...............................................................................
%...............................................................................

%...............................................................................
\begin{figure*}
\begin{center}
\includegraphics[width=\linewidth]{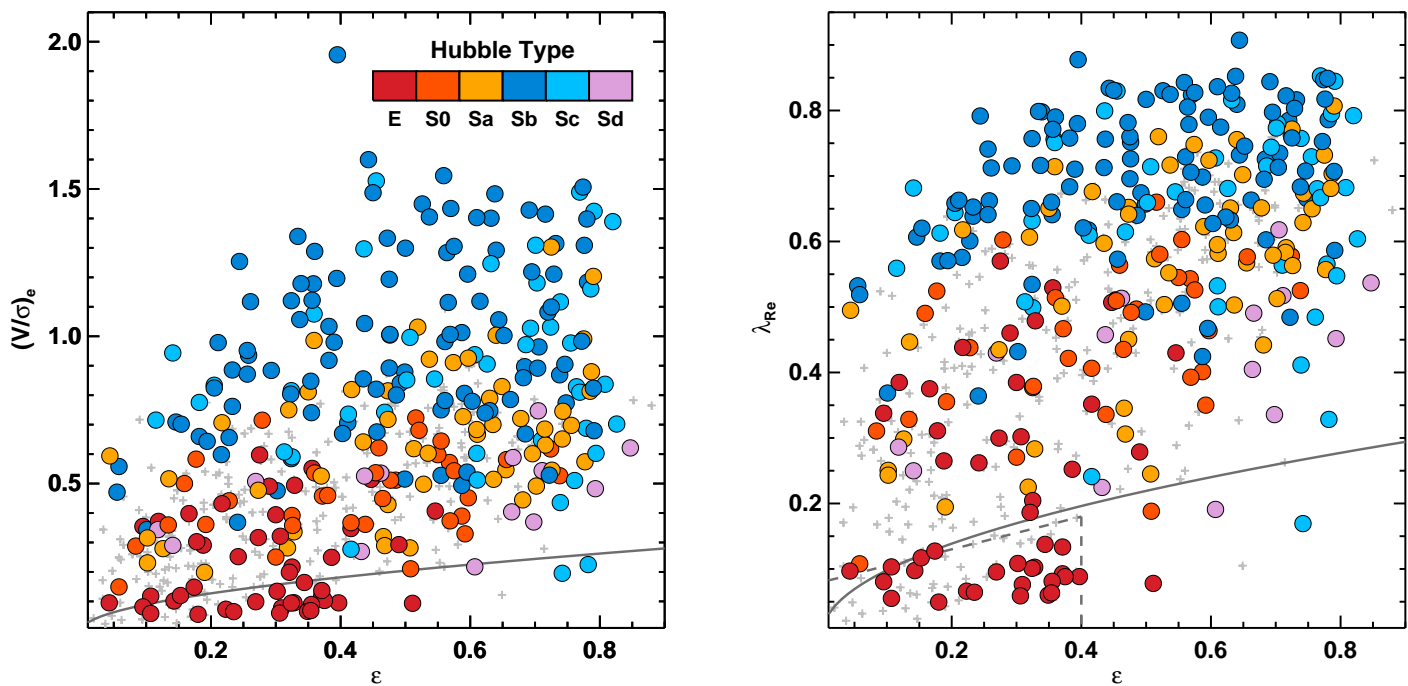}
\caption{$(V/\sigma,\eps)_e$ (left) and $(\lamRe,\eps)$ (right) relations for the 
CALIFA sample of 300 galaxies. Galaxies are color-coded with Hubble type as indicated 
by the colorbar. For reference, we plot the \atlas\ sample with gray crosses. The solid line 
demarcates the division between Slow and Fast rotators as established by \citet{emsellem_etal_2011}. 
The dashed line in the right panel marks the revised division between Slow and Fast rotators defined 
by \citet{cap16}.}
\label{fig:VSelamRe}
\end{center}
\end{figure*}
%...............................................................................

%---------------------------------------------------------------------
\subsection{Integrated radial profiles}
\label{sec:lamRproftype}
%---------------------------------------------------------------------

Figure~\ref{fig:profiles} shows the apparent stellar angular momentum \lamR\  
defined in equation~\eqref{eq:deflamR} as a function of increasing enclosed 
radius (R) along the major axis, normalized by the effective radius \Reff\ of 
each galaxy. The color represents the Hubble type of the galaxy, from 
spheroid-dominated ellipticals in red to disk-dominated spirals in blue.

The elliptical galaxies typically have the lowest \lamR\ values at 
a given (normalized) radius, even though in most cases the angular momentum does 
gently rise at larger radii. This is in line with the significant net rotation 
observed from radial velocity measurements of planetary nebulae and globular 
clusters in the outskirts of elliptical galaxies \citep[e.g.][]{sluggs17}. 
Moreover, even giant ellipticals like M87 in the Virgo Cluster that appear 
round, in deeper images do show in the outskirts significant flattening 
reflecting at least partial rotational support \citep[e.g.][]{liu05}. Additional 
evidence is found in early-type galaxies with faint spiral-like structures found at 
large radii \citep[e.g.][]{gomes16}. Our findings are consistent with dedicated 
studies of early-type galaxies reaching up to 5\,\Reff\ \citep[e.g.][]{raskutti14,boardman17}.

Rather unexpected is that the galaxies which have the \lamR\ profiles with the 
largest amplitudes are not the most disk-dominated spiral galaxies. Already in 
the inner parts, the stars in Sb galaxies have a larger apparent angular 
momentum than S0 and Sa galaxies, as anticipated from the larger disk-to-total 
fractions of Sb compared to S0/Sa galaxies. However, the \lamR\ values of Sb 
galaxies are on average also significantly higher than for Sc and Sd galaxies 
even though the latter are relatively more disk dominated. The most extreme 
cases in our sample are MCG-02-51-004 (ID: 868), NGC6301 (ID:849), and 
UGC12518 (ID: 910). See Table~\ref{tab:sample} for details.

Since \lamR, as opposed to $V/\sigma$, is normalized in 
equation~\eqref{eq:deflamR} by the sum of the squares of velocity ($V$) and 
dispersion ($\sigma$), it not only has a well-defined maximum of unity, but 
should also be nearly independent of mass --- the enclosed total mass is namely 
proportional to the second velocity moment, which after projection and 
integration along the line of sight, in turn is proportional to $V^2+\sigma^2$. 
Therefore, the difference in \lamR\ profiles between galaxies of different 
morphological type can not merely be the result of a possible difference in 
mass. 

%...............................................................................
\begin{figure*}
\begin{center}
\begin{minipage}{\linewidth}
\includegraphics[width=\linewidth]{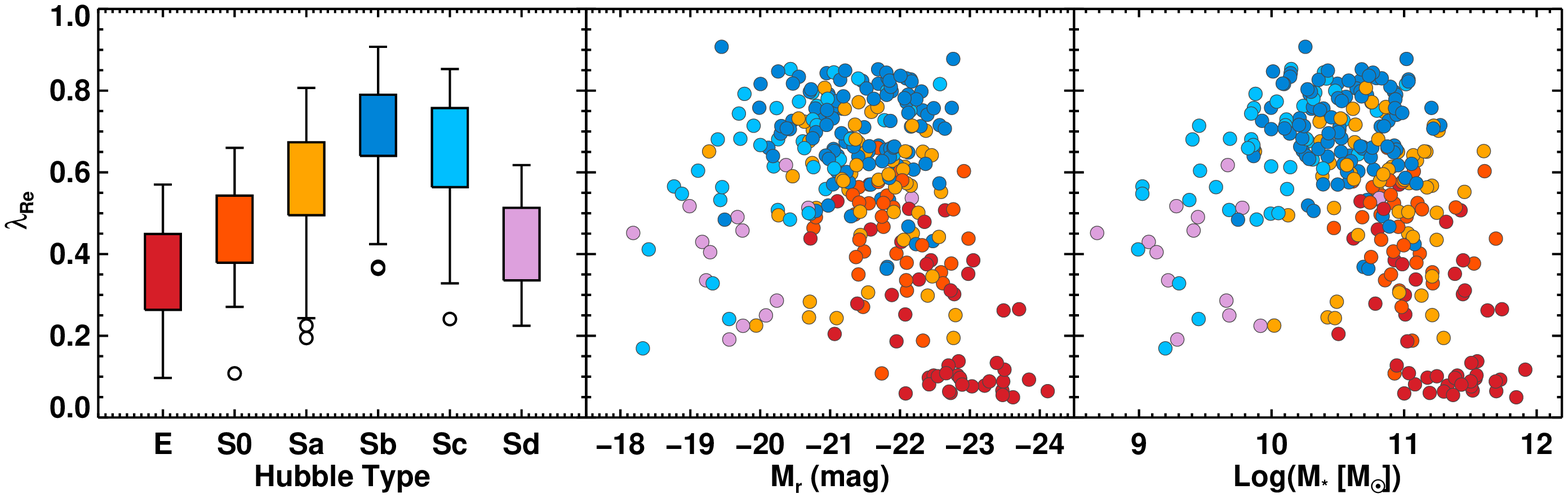}
\vspace{-5pt}
\end{minipage}
\begin{minipage}{\linewidth}
\includegraphics[width=\linewidth]{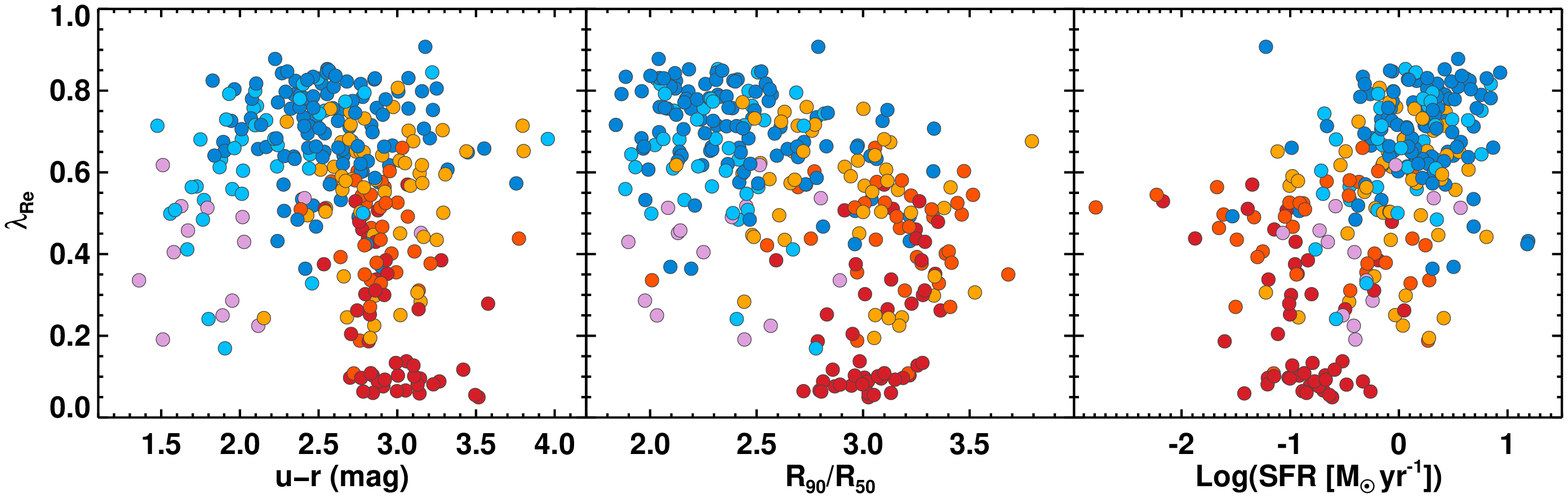}
\end{minipage}
\caption{\lamRe\ relations with global parameters for the sample of 300 CALIFA galaxies. Galaxies 
are color-coded with Hubble type. The top left panel shows a whisker plot enclosing the interquartile 
range (IQR), defined at IQR25\%-IQR75\% for galaxies of each morphological type. The 
whiskers extend out to the maximum or minimum value of the data, or to 1.5 times
IQR25\% or IQR75\% in case there is data beyond this range. Outliers are identified with small circles. 
Other panels show relations with $r-$band absolute magnitude $M_r$, total stellar mass $M_\star$, $u-r$ Petrosian 
color, concentration index R$_{90}$/R$_{50}$, and star formation rate SFR. See text for details.}
\label{fig:lamRe_correlations}
\end{center}
\end{figure*}
%...............................................................................

%=====================================================================
\section{Rotation versus pressure support: $(V/\sigma,\eps)$ and ($\lamRe,\eps$) diagrams}
\label{sec:lamRepsdiagram} 
%=====================================================================

From earlier studies of E/S0 galaxies, in particular from the \sauron\ project 
\citep{emsellem07,cappellari_etal_2007} and \atlas\ Survey 
\citep{emsellem_etal_2011}, we know that the slow-rotator and fast-rotator 
galaxies, apart from their different position in the $(\lamRe,\eps)$-diagram, do 
have other distinct properties. Slow rotators often show kpc-scale kinematically 
distinct cores (KDCs) with similarly-old ages as the rest of the stars in a 
galaxy that tends to be a quiescent, massive elliptical galaxy with a mildly 
triaxial intrinsic shape \citep[e.g][]{mcdermid_etal_2006}. Most elliptical 
galaxies and nearly all lenticular galaxies are, however, fast rotators having 
an intrinsic shape, apart from the common presence of bars, that is fully 
consistent with oblate axisymmetry and spanning a wide range in flattening. They 
show regular rotation with aligned photometric and kinematic axes even though a 
KDC is sometimes present, but typically of smaller scale than in slow rotators 
and containing stars that are on average younger than the main body. 
A similar picture is revealed by the E (red) and S0 (orange) galaxies from the 
CALIFA survey. 

Figure~\ref{fig:VSelamRe} shows both the more traditional ordered-over-random 
stellar motion $\VSe$ (left panel) and apparent stellar angular momentum \lamRe\ 
(right panel) as function of the ellipticity $\eps$. The solid curve indicates 
the demarcation line between slow-rotator and fast-rotator galaxies 
as inferred from the \atlas\ survey of elliptical (E) and lenticular (S0) galaxies. 
The CALIFA survey explores in a homogeneous way galaxies of all Hubble types, as 
indicated by the color of the symbols. The Sa and Sb galaxies show the expected 
continuation of fast-rotator E/S0 galaxies: reaching higher values of $\VSe$ and 
$\lamRe$ and having on average larger $\eps$, consistent with the increasing 
dominance of a disk with resulting increase in rotational support and flattening.
Interestingly though, the rotational support is \emph{decreasing} again with Sc 
and in particular Sd galaxies, some of which reach \lamRe\ values close or even 
below the slow-fast-rotator demarcation line. Still, they remain very different 
from slow-rotator elliptical galaxies because the spiral galaxies have much 
larger $\eps$ values and hence are intrinsically much flatter. We explore this 
behaviour in more detail in the next section (\S\ref{sec:trends}).

We choose to plot in Fig.~\ref{fig:VSelamRe} as a reference the \atlas\ sample, as they provide values 
for both $\VSe$  and \lamRe. The comparison of CALIFA with \atlas\ and other samples  
in the literature is overall good. While differences in the range of measured ellipticities are small, the 
biggest discrepacies appear in the range of \lamRe\ values. Differences with 
C16 and vdS17 are mostly on the maximum values of \lamRe\ reached. While 
our largest values are around \lamRe$\sim$0.85, the SAMI survey galaxies 
hardly go over 0.8. This is in contrast with the MaNGA sample of G18 that displays 
\lamRe\ values that often reach (and pass) the theoretical maximum of 1.0.  
As opposed to G18 galaxies, our sample lacks round, fast rotating galaxies, 
which may be due to the CALIFA sample selection that precludes the inclusion of large, 
face-on disks. Interestingly, the range of \lamRe\ values of \citet{sanchez18}, also based on 
MaNGA data, is consistent with the ones presented here. The sometimes extreme 
particularities of the beam corrections applied in G18 as opposed to \citet{sanchez18} may 
be at the heart of the large differences between the two studies on the same dataset. 
The similar effect is also true for when comparing our sample with that of vdS17. In 
this particular case, differences can be due to the particular definition the SAMI team adopted for R$_j$ in 
equation~\ref{eq:deflamR} (e.g. semi-major axis of the ellipse on which spaxel $j$ lies, instead of the 
circular projected radius to the center). This also results in a lower \lamRe\  value as compared to the 
Polar R$_j$ definition that is adopted here. Regardless of the specific details in the sample 
selection and  peculiarities in the \lamRe\ calculation of the three surveys, they are largely 
complementary.\looseness-2
   
%---------------------------------------------------------------------
\subsection{Trends with global parameters}
\label{sec:trends}
%---------------------------------------------------------------------

To investigate further the properties of galaxies of all morphological types in the 
$(\lamRe,\eps)$-diagram, we show in Fig.~\ref{fig:lamRe_correlations} the relation 
between \lamRe\ and different global parameters, color-coding galaxies according 
to their Hubble type. 

The top row in Fig.~\ref{fig:lamRe_correlations} shows the behaviour of \lamRe\ with Hubble 
type, $r-$band absolute magnitude and total stellar mass (from left to right).  Not surprisingly 
the ellipticals display a wide range of \lamRe\ values, from the lowest in the sample 
close to zero to almost 0.6. As originally observed in the \sauron\ survey, the E 
family comprises galaxies that includes both slow and fast rotators. The middle and 
right panels confirm that luminosity and mass are the best predictors for slow rotators, 
being the the dominant population at the high luminosity and mass end. Nevertheless, 
the increase of \lamRe\ with Hubble type would still hold even if slow rotators were not 
considered. This increasing trend with morphological type was already observed 
by C16 in the SAMI survey. Interestingly, though, our sample shows a maximum 
in \lamRe\ for the Sb galaxies, with decreasing values for later-types. We have used the much 
larger MaNGA sample of G18 to check this trend. While there is indeed a turning 
point at similar stellar masses, this is much milder than observed in our CALIFA sample at low 
masses. We attribute the difference to the peculiarities of our sample, which is not complete 
for the low luminosity (and thus mass) end (see FLV17 for more details).

The bottom row shows the relation between \lamRe\ with $u$\,$-$\,$r$ Petrosian color, concentration 
index (R$_{90}$/R$_{50}$, measured as the ratio of 90 and 50 percentile Petrosian radius), and 
star-formation rates derived from H$\alpha$ emission line fluxes in \citet{sanchez17}. 
Our sample of E/S0/Sa galaxies show a well-defined vertical sequence around the same red colour, while
later types display bluer colors. The middle panel shows that, slow-rotators aside, the concentration
index is clearly related to \lamRe\, with larger values reached by the Sb types. This is expected 
as light concentration is a proxy for the bulge($+$bar)-to-total ratio, which in turn is the main 
driver behind the Hubble morphological classification \citep[e.g.][]{Shimasaku01,Strateva01}.
Still, there appears to be some scatter in the relation, consistent with that shown already in 
C16. The  family of Sb galaxies also appear to be the currently higher star-forming 
systems (right panel). Similarly, going from spiral to elliptical galaxies, the SFR decreases, so it 
also unsurprising to find the trend of decreasing \lamR\ with lower SFR.\looseness-1

The \lamRe\ values found for the galaxies in our sample confirm the predominance of Slow 
rotators in high-mass, high-luminosity systems. We estimate an overall fraction of 28\% of 
Slow rotators with stellar masses above 10$^{11}$\,\Msun, based on the \citealt{emsellem_etal_2011} 
definition. This number sits in the middle of the wide range of predictions provided by the latest surveys
\citep[e.g.][]{emsellem_etal_2011, deugenio13, fogarty14, veale17a, vdsande17b} which 
display values between 15\% and 80\% for masses above 10$^{11}$\,\Msun. Our lower value 
is likely due to the fact that the CALIFA survey is complete only up to 10$^{11.44}$\,\Msun\  
\citep[see][for details]{Walcher_etal_2014}.\looseness-2

Despite the limited number of galaxies compared to other surveys, our sample shows two 
areas with interesting results: (i) the low \lamRe\ values for the late-type spirals, and (ii) the 
particular properties of the fastest rotators.\smallskip

{\bf Low \lamRe\ spirals:} we have investigated the reasons for the surprisingly low \lamRe\ values 
observed in the latest-type galaxies and found two potential explanations. There is a group of Sc/Sd 
galaxies with \lamRe\ values below 0.35. We have checked and these are both irregular 
or fairly face-on systems. This naturally explains their unsual location in the ($\lamRe,\eps$) 
diagram. The same feature was found by \citet{graham18} in their much larger sample 
of galaxies. The remaining group of Sc/Sd galaxies with \lamRe\ values between 0.35 and 
0.6 are typically edge-on systems. We have explored whether they present large extinction values, 
as dust obscuration could prevent the full integration of the stellar kinematics along the line-of-sight 
and thus led to lower rotation amplitudes. Displaying the bluest colors of the entire sample, this 
option does not seem to be likely. This is confirmed by the much more detailed study of the 
extinction in the CALIFA galaxies by \citet{rosa15}. Note, however, that simulations suggest that 
observational estimates could be understimated for this kind of systems \citep[see][for details]{ibarra19}. 
In addition, we have also checked that those galaxies display velocity dispersions well above the 
point where the limits in the CALIFA spectral resolution are an issue (see Fig. 9 in FLV17).

The large observed $\eps$ values for those Sc/Sd galaxies imply that we need a way 
to keep their dynamically hot stellar disk geometrically thin. We note that they contain 
small bulges (as observed by their low concentration values), and also are not the 
highest star-forming galaxies. We postulate that the presence of a relatively large dark 
matter halo provides an additional vertical force to keep the disk geometrically thin while being 
dynamically heated. Our initial assessment, based on dynamical models of our sample 
\citep{zhu18}, suggests an enclosed mass that is up to a factor ten larger than the 
estimated baryonic (stellar plus gas) mass already within the half-light radius. A 
preliminary confirmation of this was presented in Fig.~3 of \citet{fb15}. This is in line with previous 
results in the literature presenting evidence of \textit{thicker} thin disks in late-type spirals 
\citep[e.g.][]{yoachim06,comeron11}.\smallskip

%...............................................................................
\begin{figure}
\begin{center}
\begin{minipage}{\linewidth}
\includegraphics[width=\linewidth]{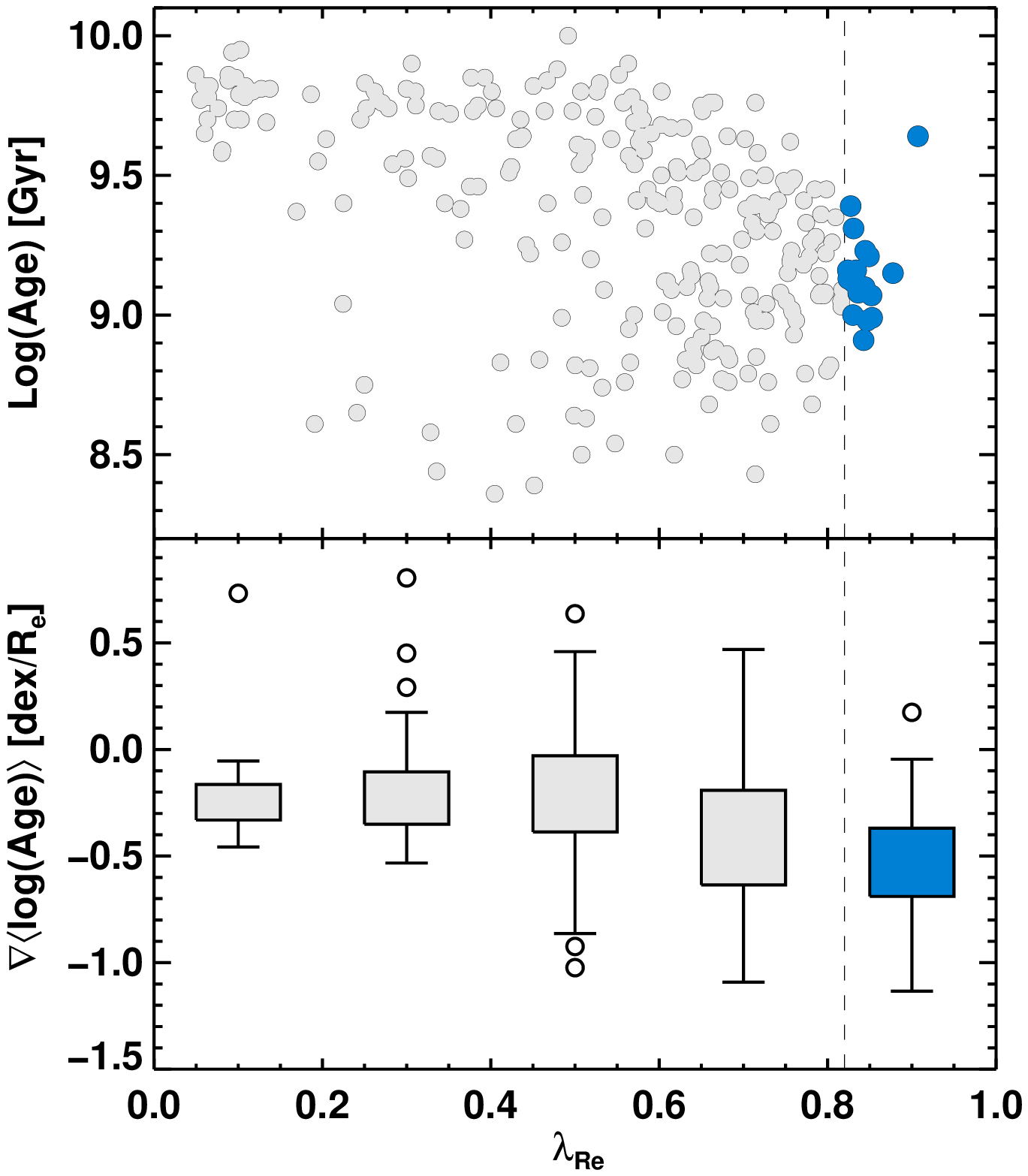}
\vspace{-5pt}
\end{minipage}
\begin{minipage}{\linewidth}
\includegraphics[width=\linewidth]{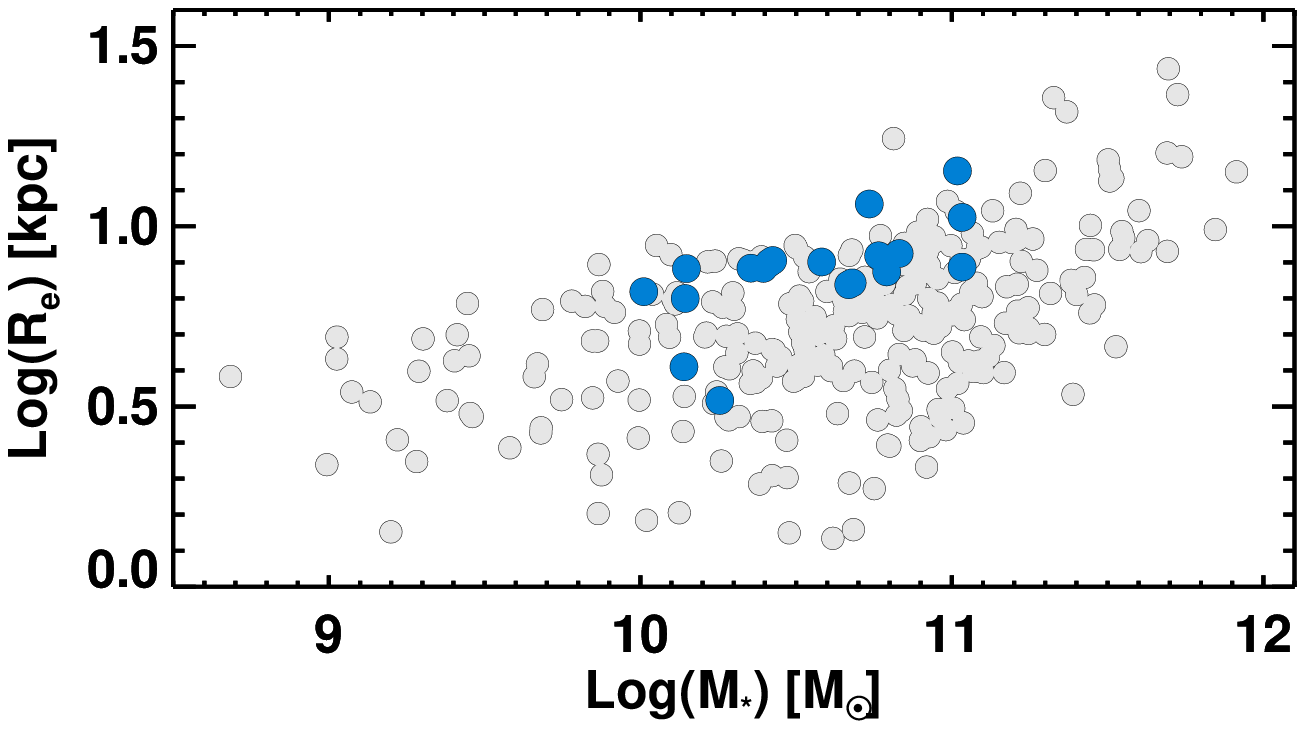}
\end{minipage}
\caption{(Top and middle panels) \lamRe\ relations with mean age within \Reff, mean stellar 
age gradient within \Reff\  for our sample of CALIFA galaxies. (Bottom panel) Stellar mass --- size 
relation. The whisker plot in the middle panel was computed in the same way as in Fig.~\ref{fig:lamRe_correlations}. 
Galaxies above the 95\% percentile of \lamRe\ distribution of the sample are marked in blue. 
The dashed line in the top and middle panels marks that percentile. }
\label{fig:superfast}
\end{center}
\end{figure}
%...............................................................................

\noindent{\bf Fastest rotators:}  We have identified a group of 19 galaxies with 
\lamRe\,$\ge$\,0.82 (i.e. the top 5\% of the distribution). They are mostly Sb/Sc 
galaxies. In Fig.~\ref{fig:lamRe_correlations} they happen to have intermediate 
absolute magnitudes, masses, and colors. They are not special in any of those three 
parameters with respect to other galaxies with lower \lamRe\ values. They are peculiar 
in that they are the highest star-forming galaxies with some of the smallest 
bulges, as probed by the concentration index\footnote{This is confirmed by the detailed 
bulge/disk photometric decomposition of \citet{jairo17} for the limited subset of CALIFA 
galaxies in common with the sample presented here.}. We have identified at least three  
other properties that make these galaxies unique. As shown in Fig.~\ref{fig:superfast}, 
they tend to have the largest sizes at a given stellar mass, display some of the 
strongest average luminosity-weighted inner age gradients measured by \citet{rgb17}, 
and also appear to have rather homogeneous mean stellar population ages within \Reff\ 
of  about 1\,Gyr \citep{rosa15}. We inspected for any dependence with environment,
either local or global, and found no significant trends.

The relative difference between the observed \lamRe\ values of this group of galaxies 
(see top, left panel of Fig.~\ref{fig:lamRe_correlations}) with respect to the S0 population, 
raises the question whether they could fade into lenticular galaxies. Decades after the 
discovery of the morphology-density relation \citep{dressler87}, the interest in this topic 
has been revived by the recent results from different groups \citep[e.g.][]{bedregal06, 
laurikainen10, kormendy12, brough17,greene17} confirming the initial result, but recasting the observed 
phenomenon from the Slow/Fast rotator perspective \citep[e.g.][]{cappellari_etal_2011b}.
We refer the reader to the extended review on the topic presented in \citet{cap16}.
At first sight, based purely on \lamRe\, our results suggest that the transformatiom 
between Sa galaxies into S0s is possible. Note, however, that in a fading scenario, both 
stellar mass and \lamRe\ are expected to be conserved. This seems harder for Sb and 
Sc galaxies, for which the difference with respect to the lenticulars in \lamRe\ is significant. 
For Sd galaxies, even though \lamRe\ values are consistent with those of lenticulars, their 
light concentrations are much lower and thus it seems unlikely they will fade into lenticulars
with typically much larger bulge-to-disk ratios. Furthermore, the amount of mass in gas for 
these late-type galaxies can be up to 50\% of their baryonic total mass \citep[e.g.][]{papastergis12}, 
making it very difficult to turn all that matter into stars by fading within a Hubble time without 
substatially increasing the total mass budget of the system. Pre-processing in groups, with 
tidal interactions and/or major mergers seem to be more likely mechanisms \citep[e.g.][]{querejeta15}. 

%...............................................................................
\begin{figure*}
\begin{center}
\includegraphics[angle=0,width=\linewidth]{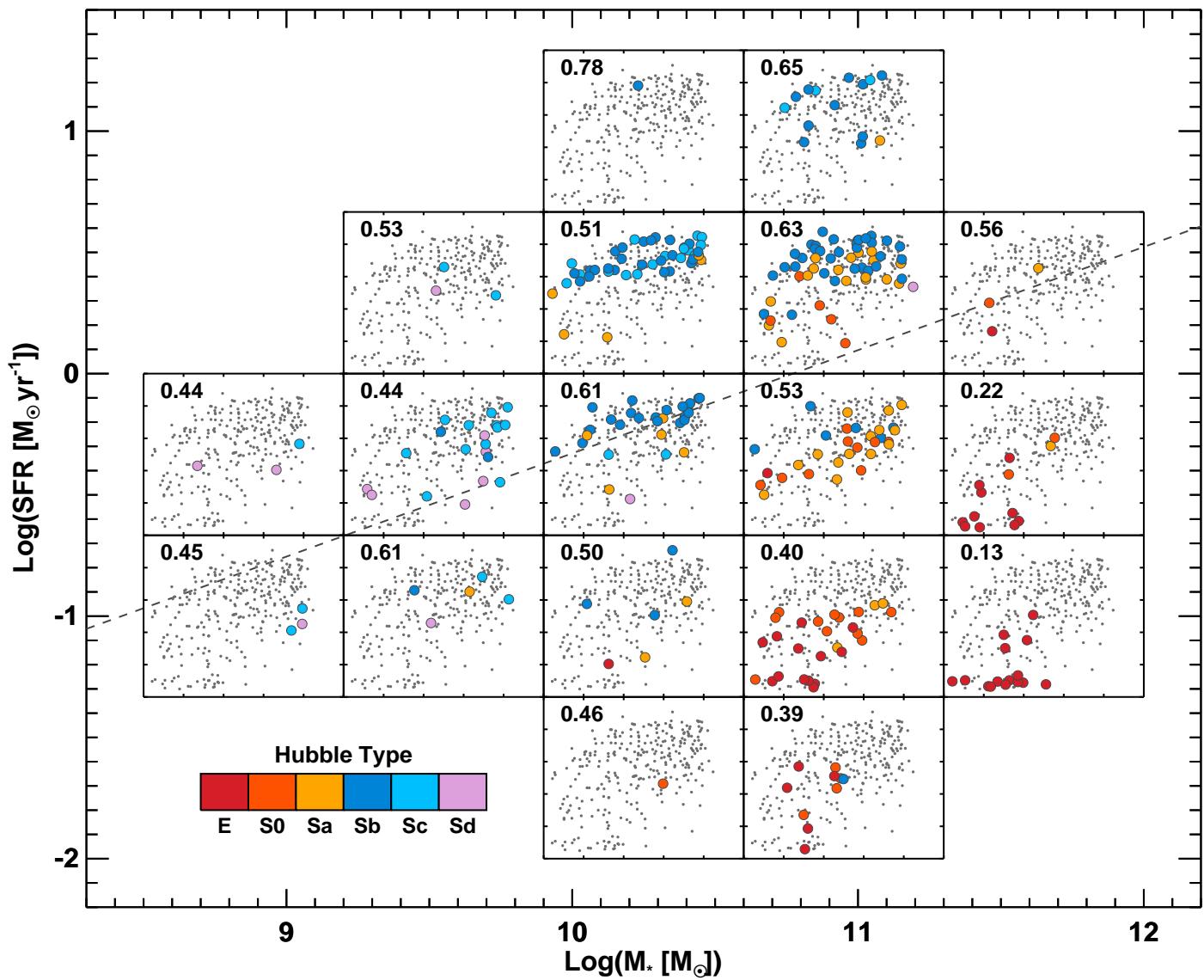}
\caption{The \lamRe\ - $\epsilon$ relation as a function of location in the plane of  star-formation 
rate versus total stellar mass $\Mstar$ for the CALIFA galaxies. Each panel shows the complete 
sample highlighting in color the ones belonging to each specific SFR-$\Mstar$ bin. The number in 
each subpanel gives the V$_{\rm max}^{-1}$-weighted \lamRe\ average for the highlighted 
galaxies in that bin. The dark gray dashed line in the SFR versus $\Mstar$ on the right marks 
the division between the main star-forming sequence and quiescent galaxies from \citet{renzini15}.\looseness-2}
\label{fig:cmd}
\end{center}
\end{figure*}
%...............................................................................

%---------------------------------------------------------------------
\subsection{Trends in the SFR-$\Mstar$ diagram}
\label{sec:trends_cmd}
%---------------------------------------------------------------------

An interesting way of looking at variations of \lamRe\ is through the 
extensively studied star formation rate\,--\,stellar mass relation 
\citep[e.g.][]{croton06,cortese19}, shown in Fig.~\ref{fig:cmd} 
for our sample. We have divided the diagram in bins of SFR and stellar mass. 
Each subpanel presents the ($\lamRe,\eps$) relation for that bin, showing in 
gray all CALIFA galaxies and highlighting in colour those belonging to that bin 
with their Hubble type. The number on the top-left corner of each inset gives the 
V$_{\rm max}^{-1}$-weighted \lamRe\ average for the highlighted galaxies in 
that bin. 

The figure shows that the main star-forming sequence is made of galaxies with 
increasing \lamRe\ as both the SFR and stellar mass grow. The high mass and high 
star-forming region is populated by Sa/Sb galaxies mostly, whereas the 
low SFR and stellar mass ends are dominated by later-type galaxies (Sc/Sd).  
As already highlighted by numerous studies, galaxies departing from the main star-forming 
sequence are mostly early-type Es and S0s \citep[e.g.][]{schawinski14}. It appears that the dynamically 
coldest disks are found in the most massive and more actively star-forming systems \citep[e.g.][]{catalan17,jairo19}. 
The trends with \lamRe\ observed here are supported by similar analysis 
with the EAGLE cosmological simulation (Walo-Mart\'in et al., in preparation).

%=====================================================================
\section{Conclusions}
\label{sec:conclusions} 
%=====================================================================

This paper presents the CALIFA view on the stellar angular momentum distribution 
for a sample of 300 galaxies across the Hubble sequence. Our dataset allows us 
to study the distribution of \lamR\ for different apertures (0.5\,\Reff, \Reff, 
2\,\Reff) and provides the relationship between them, including conversions to 
\lamR\ computed with a long-slit along the major axis of the galaxies. Our sample 
also helps us to investigate the relationship between \lamR\ and different global 
properties of galaxies (e.g. Hubble type, absolute magnitude, $u-r$ color, 
concentration index, stellar mass, and star formation rate). 

In addition, we analyze the distribution of galaxies in the classical 
$(V/\sigma,\eps)$ and ($\lamRe,\eps$) diagrams, often used to study the level 
of rotation over pressure support in galaxies. Our results for early-type (E and 
S0) galaxies are consistent with previous studies in the literature for the same 
kind of galaxies. The extension to later-types (Sa to Sd) provided by our sample 
presents two interesting results. On one side, we find a maximum \lamRe\ of 
around $\sim$0.85 for large, relatively massive and highly star-forming galaxies 
(typically Sb systems). On the other hand, rather unexpectedly, we observe 
relatively low \lamRe\ values for low-mass Sc/Sd systems. We will exploit these 
two areas in forthcoming papers to discuss the nature of S0 galaxies, and to 
investigate the dark matter content of low mass systems respectively.\looseness-2

The results presented here with the CALIFA sample in terms of the stellar 
angular momentum distribution of galaxies are just the tip of the iceberg of 
possibilities to extend our understanding of galaxy formation and evolution. 
Ongoing large surveys have already started to exploit this information in 
different areas (e.g. vdS17), with a boost in this field with the measurement 
of \lamRe\ for thousands of galaxies provided by the MaNGA survey team 
(e.g. G18). Complementary, the first studies relating the radial dependence of 
\lamR\ to the evolution of galaxies are appearing in the literature \citep[e.g.][]{graham17}.
In the absence of high-quality observations of stellar kinematics for 
substantial samples of high-redshift galaxies (e.g. $z>1.0$), (cosmological) 
numerical simulations will allow us to explore the evolution of angular momentum with time 
\citep[e.g.][]{lagos18,schulze18,pillepich19,vds19}.

%=====================================================================
% ACKNOWLEDGEMENTS
%=====================================================================
\begin{acknowledgements}
We would like to thank the anonymous referee for constructive comments
that helped improving some aspects of the original manuscript.
This study makes use of the data provided by the Calar Alto Legacy
Integral Field Area ({\sc CALIFA}) survey (http://www.califa.caha.es).
Based on observations collected at the Centro Astron\`{o}mico Hispano
Alem\'{a}n (CAHA) at Calar Alto, operated jointly by the
Max-Planck-Institut f\"{u}r Astronomie and the Instituto de
Astrofisica de Andalucia (CSIC). {\sc CALIFA} is the first legacy
survey being performed at Calar Alto. The {\sc CALIFA} collaboration
would like to thank the IAA-CSIC and MPIA-MPG as major partners of the
observatory, and CAHA itself, for the unique access to telescope time
and support in manpower and infrastructures.  The {\sc CALIFA}
collaboration thanks also the CAHA staff for the dedication to this
project.\medskip\newline
{\it Funding and financial support acknowledgements}: JF-B. from grant 
AYA2016-77237-C3-1-P from the Spanish Ministry of Economy and Competitiveness 
(MINECO); GvdV acknowledges funding from the European Research Council (ERC) 
under the European Union's Horizon 2020 research and innovation programme 
undergrant agreement no.~724857 (Consolidator Grant ArcheoDyn). 
BG-L. acknowledge support from the State Research Agency (AEI) of the Spanish Ministry 
of Science, Innovation and Universities (MCIU) and the European Regional Development Fund 
(FEDER) under grant with reference AYA2015-68217-P. SFS is grateful for the support of 
a CONACYT grant CB-285080 and FC-2016-01-1916, and funding from the 
PAPIIT-DGAPA-IA101217 (UNAM) project.
LZ acknowledges support from  Shanghai  Astronomical  Observatory,  Chinese  
Academy  of Sciences under grant no. Y895201009. L.G. was supported in part by 
the US National Science Foundation under Grant AST-1311862. RGD  from AYA2016-77846-P,  
AYA2014-57490-P,  AYA2010-15081, and Junta de Andaluc\'i a FQ1580. IM from grants 
AYA2013-42227-P and AYA2016-76682-C3-1-P. RGB, RMGD, IM, and EP acknowledge financial 
support from the State Agency for Research of the Spanish MCIU through the "Center of Excellence 
Severo Ochoa" award to the Instituto de Astrof\'isica de Andaluc\'ia (SEV-2017-0709). 
\end{acknowledgements}

%=====================================================================
% REFERENCES
%=====================================================================

\bibliographystyle{aa}
\bibliography{califaam}

%=====================================================================
% APPENDICES
%=====================================================================
\appendix

%=====================================================================
\section{Deprojection of \lamRe}
\label{app:deproj}
%=====================================================================
For an oblate galaxy, viewed at inclination $i$, we obtain
\begin{equation}
  \label{eq:deprojeps}
  \epsintr =  1 - \frac{\sqrt{(1-\eps)^2 - \cos^2i}}{\sin i},
\end{equation}
as the deprojection of the observed ellipticity $\eps$ to the intrinsic 
ellipticity $\epsintr$ when the galaxy would be viewed edge-on $(i=90^\circ)$.
Defining the global anisotropy parameter as 
$\delta \equiv 1 - 2\sigma_z^2/(\sigma_R^2+\sigma_\phi^2)$, we infer as shown by \citet{binney05}
\begin{equation}
  \label{eq:deprojVS}
  \VSintr = \frac{\sqrt{1 - \delta \cos^2i}}{\sin i}  \,  (V/\sigma),
\end{equation}
for the deprojection of the observed ordered-over-random motion to its edge-on 
value. Next, inserting this into the approximate relation between \lamR\ and 
$V/\sigma$ (see Eq.~B1 of \citealt{emsellem_etal_2011}), we find
\begin{equation}
  \label{eq:deprojlamR}
  \lamRintr \simeq \frac{\sqrt{1 - \delta\cos^2i}}{\sin i} \frac{\lambda_R}{ \sqrt{ 1 + (1 - \delta) \cot^2i \, \lambda_R^2} },
\end{equation}
as the approximate deprojection of the observed apparent stellar angular 
momentum \lamR\ to its edge-on value $\lamRintr$.

The inclination of a galaxy can be inferred directly from observations only in 
special cases, for example when a intrinsically thin \emph{and} circular disk 
(in cold gas or corresponding dust) is present, so that its inclination follows 
directly from the observed ellipticity because $1-\eps = \cos i$. In general, 
even if disks are close to axisymmetric, they have a non-negligible intrinsic 
flattening $q$ which, moreover, will vary from galaxy-to-galaxy.

If we assume that each galaxy is drawn from a group of galaxies with 
intrinsic shape distribution $f(q)$, the probability of viewing the 
galaxy at inclination $i$ is given by its observed ellipticity $\eps$ as
\begin{equation}
  \label{eq:probigiveneps}
  f(i|\eps) = \frac{f(q) (1-\eps)}
  {\sqrt{\sin^2i - \eps(2-\eps)}},
\end{equation}
for $0 \le \cos i < 1-\eps$, and zero otherwise.

We adopt for $f(q)$ a Gaussian distribution with mean and standard deviation 
$(\mu_q,\sigma_q)$, even though it is well known that this cannot fit the 
observed ellipticity distribution of a complete (and hence randomly inclined) 
sample of galaxies. For example, \cite{lambas92} introduce even for 
spiral galaxies an additional Gaussian distribution in the intermediate-to-long 
axis ratio $p$ with mean around the oblate case of $p=1$, but also non-zero 
dispersion to fit the tail toward rounder galaxies. However, the effect on the 
inferred (median) inclination is small, even for the mildly triaxial 
slow-rotator elliptical galaxies, so that we assume an oblate intrinsic shape 
for all galaxies. Even more so, it turns out that the Gaussian distribution with 
$(\mu_q,\sigma_q)=(0.25,0.12)$ inferred for 13,482 spirals by \cite{lambas92} 
is nearly identical to the Gaussian distribution with 
$(\mu_q,\sigma_q)=(0.25,0.14)$ inferred the fast-rotator E/S0 galaxies from the 
\atlas\ survey by \cite{weijmans14}. Henceforth, for all fast-rotator galaxies 
we adopt the latter Gaussian intrinsic shape distribution, whereas \cite{weijmans14}
find that the intrinsic shape distribution of the slow-rotator galaxies 
is well described by a Gaussian with $(\mu_q,\sigma_q)=(0.63,0.09)$.

Based on axisymmetric dynamical models of 24 E/S0 galaxies, 
\citet{cappellari_etal_2007} find that their velocity distribution is close to 
isotropic in the equatorial plane ($\sigma_R \sim \sigma_\phi$) and the 
remaining anisotropy in the meridional plane ($\delta \simeq \beta \equiv 1 - 
\sigma_z^2/\sigma_R^2$) is approximately linearly correlated with intrinsic 
ellipticity $\epsintr$. Based on this correlation, adopting a Gaussian 
distribution in $\delta$ with mean $\mu_\delta = 0.5 \, \epsintr$ and standard 
deviation $\sigma_\delta = 0.1$ for $0 \le \delta \le 0.8~\epsintr$ and zero 
elsewhere, \citet{emsellem_etal_2011} can explain the observed 
$(V/\sigma,\eps)$-diagram of the complete sample of \atlas\ fast-rotator E/S0 
galaxies. 

We followed the procedure above for each CALIFA galaxy to approximate the 
observed \lamRe\ values into intrinsic $\lamReintr$ values at an edge-on view. 
First, adopting the above fast-rotator or slow-rotator Gaussian intrinsic shape 
distribution $f(q)$ to obtain the average inclination $\iav$ from the median of 
the corresponding distribution in inclinations given by 
equation~\eqref{eq:probigiveneps}. Second, inserting $\iav$ and the observed 
ellipticity in $\eps$ into equation~\eqref{eq:deprojeps}, which provides the 
approximate intrinsic ellipticity $\epsintr$. Finally, adopting the above Gaussian 
distribution in the global anisotropy $\sigma_\delta$, equation~\eqref{eq:deprojlamR} 
provides the approximate deprojection to the intrinsic stellar angular momentum 
$\lamReintr$ within the effective radius \Reff.

%============================================================================================
\clearpage
\renewcommand\thetable{1}

\begin{deluxetable}{lccccccccccccccc}
\tablewidth{0pt}
\setlength{\tabcolsep}{3.00000pt}
\tablecaption{Stellar angular momentum properties of the CALIFA stellar kinematics sample}
\tablehead{\colhead{ID} & \colhead{$\epsilon$} & \colhead{PA} & \colhead{incl.} & \colhead{R$_{\rm eff}$} & \colhead{Type} & \colhead{M$_*$} & \colhead{M$_r$} & \colhead{C$_{90/50}$} & \colhead{$u-r$} & \colhead{SFR} & \colhead{$\lambda_{\rm 0.5\,Re}$} & \colhead{$\lambda_{\rm Re}$} & \colhead{$\lambda_{\rm 2\,Re}$} & \colhead{$\lamReintr$} & \colhead{$(V/\sigma)_{\rm e}$}
\\
\colhead{~} & \colhead{~} & \colhead{(deg)} & \colhead{(deg)} & \colhead{(arcsec)} & \colhead{~} & \colhead{(10$^{10}$\,\Msun)} & \colhead{(mag)} & \colhead{~} & \colhead{(mag)} & \colhead{(\Msun\,yr$^{-1}$)} & \colhead{~} & \colhead{~} & \colhead{~} & \colhead{~} & \colhead{~}
}
\startdata
  1 & 0.685 &   3.4 &  75.8 &  16 & Sb &  4.52 & -21.10 &   3.33 &   2.78 &   0.98 &   0.51 &   0.60 &   0.74 &   0.61 &   0.67 \\
  2 & 0.526 &  44.1 &  65.8 &  16 & Sbc &  6.78 & -22.09 &   2.06 &   2.86 &   6.43 &   0.71 &   0.83 &    --- &   0.84 &   1.45 \\
  3 & 0.412 & 105.3 &  56.8 &  23 & Sc &  2.45 & -21.06 &   2.11 &   2.44 &   2.18 &   0.52 &   0.61 &    --- &   0.64 &   0.74 \\
  4 & 0.302 & 173.3 &  64.6 &  17 & E1 & 10.86 & -22.66 &   3.10 &   2.83 &   0.13 &   0.07 &   0.11 &    --- &   0.12 &   0.13 \\
  5 & 0.559 &  75.1 &  67.8 &  23 & Sbc &  1.39 & -20.95 &   2.12 &   2.27 &   1.41 &   0.75 &   0.84 &    --- &   0.85 &   1.55 \\
  6 & 0.370 & 143.3 &  53.6 &  11 & Sab & 17.62 & -22.26 &   3.22 &   3.29 &   0.72 &   0.39 &   0.50 &   0.64 &   0.54 &   0.53 \\
  7 & 0.611 &  18.0 &  71.8 &  10 & Sab & 10.05 & -21.69 &   2.68 &   2.86 &   1.33 &   0.37 &   0.58 &    --- &   0.59 &   0.67 \\
  8 & 0.324 & 107.6 &  50.0 &  12 & Sbc &  6.31 & -21.73 &   3.04 &   2.39 &   4.87 &   0.44 &   0.53 &   0.64 &   0.58 &   0.58 \\
  9 & 0.302 & 177.5 &  48.1 &  17 & Sb & 10.96 & -22.47 &   3.22 &   2.24 &  15.50 &   0.45 &   0.43 &    --- &   0.48 &   0.48 \\
 10 & 0.476 &  24.6 &  61.9 &  21 & Sb &  7.87 & -22.33 &   2.49 &   2.64 &   2.47 &   0.61 &   0.75 &    --- &   0.77 &   1.00 \\
 12 & 0.755 &  96.8 &  79.8 &  20 & Sc &  1.29 & -20.75 &   2.22 &   1.91 &   3.09 &   0.64 &   0.73 &    --- &   0.73 &   0.98 \\
 13 & 0.561 & 171.2 &  68.6 &  19 & Sb &  2.34 & -20.88 &   2.24 &   2.48 &   0.69 &   0.55 &   0.73 &    --- &   0.74 &   0.98 \\
 14 & 0.462 &   7.5 &  60.9 &  20 & Sd &  0.60 & -20.69 &   2.08 &   1.79 &   3.69 &   0.41 &   0.51 &    --- &   0.53 &   0.54 \\
 16 & 0.468 &  53.4 &  61.5 &  20 & Scd &  0.99 & -20.19 &   2.23 &   2.24 &   1.19 &   0.51 &   0.61 &    --- &   0.63 &   0.74 \\
 17 & 0.490 & 149.8 &  63.2 &  18 & E4 &  6.07 & -21.39 &   3.23 &   3.58 &   0.10 &   0.25 &   0.28 &    --- &   0.29 &   0.29 \\
 18 & 0.143 & 167.6 &  41.8 &  15 & E1 & 15.00 & -22.41 &   3.01 &   2.70 &   0.10 &   0.06 &   0.10 &    --- &   0.13 &   0.10 \\
 20 & 0.473 &  49.8 &  61.6 &  22 & Sa & 10.72 & -22.18 &   2.98 &   3.02 &   0.34 &   0.33 &   0.45 &    --- &   0.47 &   0.43 \\
 22 & 0.473 &  90.7 &  61.7 &  34 & Sab & 39.90 & -21.87 &   2.69 &   3.80 &   1.82 &   0.63 &   0.65 &    --- &   0.67 &   0.81 \\
 23 & 0.055 &  32.0 &  20.7 &  26 & Sb &  5.26 & -21.84 &   1.98 &   2.64 &   0.69 &   0.29 &   0.53 &    --- &   0.80 &   0.47 \\
 24 & 0.417 &   8.2 &  57.4 &  13 & Sab &  2.34 & -20.70 &   3.79 &   3.10 &    --- &   0.56 &   0.68 &   0.86 &   0.70 &   0.82 \\
 25 & 0.339 & 167.1 &  51.1 &  28 & Sb &  8.36 & -22.31 &   2.18 &   3.01 &   2.35 &   0.64 &   0.80 &    --- &   0.83 &   1.18 \\
 26 & 0.574 & 168.7 &  68.9 &  22 & Sab &  7.05 & -21.59 &   2.63 &   2.70 &   2.69 &   0.56 &   0.75 &    --- &   0.76 &   0.91 \\
 27 & 0.711 &  25.8 &  77.3 &  20 & Sd &  0.19 & -18.99 &   2.45 &   1.63 &   0.26 &   0.34 &   0.52 &    --- &   0.52 &   0.54 \\
 28 & 0.261 &  50.5 &  44.5 &  18 & Sbc &  6.65 & -22.16 &   2.44 &   2.44 &   4.87 &   0.62 &   0.71 &    --- &   0.77 &   1.12 \\
 29 & 0.742 & 112.9 &  79.0 &  23 & Sa & 12.50 & -21.90 &   2.98 &   3.02 &    --- &   0.45 &   0.63 &    --- &   0.63 &   0.65 \\
 30 & 0.324 & 175.3 &  50.0 &  15 & Sc &  2.04 & -21.14 &   2.42 &   1.88 &   2.18 &   0.50 &   0.63 &   0.66 &   0.68 &   0.82 \\
 31 & 0.205 &  64.3 &  39.1 &  20 & Sc &  4.50 & -21.91 &   1.91 &   2.45 &   5.75 &   0.54 &   0.64 &    --- &   0.73 &   0.83 \\
 32 & 0.743 &  22.7 &  79.0 &  12 & Sab &  4.24 & -21.41 &   3.13 &   2.64 &    --- &   0.55 &   0.67 &   0.82 &   0.68 &   0.75 \\
 33 & 0.359 &  88.9 &  52.7 &  21 & Sc &  6.22 & -22.15 &   2.11 &   2.54 &   5.76 &   0.64 &   0.75 &    --- &   0.78 &   1.08 \\
 35 & 0.275 &  35.4 &  45.7 &  15 & E7 &  9.16 & -21.56 &   3.10 &   3.06 &   0.04 &   0.43 &   0.57 &    --- &   0.63 &   0.60 \\
 36 & 0.777 &  15.8 &  80.6 &   6 & Sa &  6.22 & -21.26 &   3.16 &   2.65 &   3.31 &   0.43 &   0.56 &   0.66 &   0.56 &   0.57 \\
 37 & 0.724 & 171.5 &  78.3 &  11 & S0a &  4.09 & -21.20 &    --- &   2.58 &   0.22 &   0.39 &   0.58 &   0.73 &   0.58 &   0.62 \\
 38 & 0.126 &  74.9 &  30.4 &  28 & Sa & 13.52 & -22.40 &   3.33 &   2.86 &   1.16 &   0.25 &   0.30 &    --- &   0.42 &   0.28 \\
 39 & 0.739 & 158.7 &  79.1 &  23 & Scd &  0.74 & -20.23 &   2.25 &   2.09 &   0.67 &   0.62 &   0.76 &    --- &   0.76 &   1.12 \\
 40 & 0.808 &  23.6 &  82.2 &  20 & Scd &  0.49 & -19.72 &   2.31 &   2.22 &   0.64 &   0.55 &   0.68 &    --- &   0.68 &   0.84 \\
 41 & 0.774 &  54.5 &  80.4 &  17 & Sbc &  1.03 & -20.26 &   2.52 &   2.30 &   0.87 &   0.81 &   0.85 &    --- &   0.85 &   1.51 \\
 42 & 0.500 & 128.4 &  64.0 &  21 & Sbc &  3.14 & -21.69 &    --- &   2.52 &   2.25 &   0.69 &   0.82 &    --- &   0.83 &   1.30 \\
 43 & 0.355 &  15.6 &  52.7 &  13 & Sb &  3.88 & -21.11 &   2.59 &   2.66 &   3.17 &   0.50 &   0.64 &    --- &   0.68 &   0.74 \\
 44 & 0.328 &  72.2 &  67.4 &  21 & E5 & 25.18 & -22.48 &   2.96 &   3.00 &   0.12 &   0.05 &   0.10 &    --- &   0.11 &   0.10 \\
 45 & 0.455 &  32.7 &  60.2 &  19 & Scd &  2.59 & -21.40 &   2.03 &   2.49 &   4.14 &   0.78 &   0.83 &    --- &   0.84 &   1.53 \\
 46 & 0.597 &  44.2 &  70.5 &   8 & S0 &  2.95 & -20.76 &   3.17 &   2.83 &   0.02 &   0.37 &   0.46 &   0.59 &   0.47 &   0.45 \\
 47 & 0.485 &  24.2 &  62.6 &  10 & S0 &  6.64 & -21.35 &   3.46 &   2.79 &   0.02 &   0.43 &   0.50 &   0.52 &   0.51 &   0.51 \\
 49 & 0.640 &  30.3 &  73.8 &  12 & Sa &  4.11 & -21.21 &   3.00 &   2.57 &   1.39 &   0.61 &   0.76 &   0.83 &   0.76 &   1.00 \\
 50 & 0.516 &  57.7 &  65.1 &  12 & S0 &  7.48 & -21.68 &   3.06 &   3.03 &   0.46 &   0.53 &   0.66 &   0.74 &   0.67 &   0.72 \\
 51 & 0.095 &  13.9 &  26.6 &  12 & E4 & 12.25 & -22.27 &   3.13 &   2.87 &   0.06 &   0.28 &   0.34 &   0.41 &   0.52 &   0.35 \\
 52 & 0.564 & 137.5 &  68.2 &  19 & Sbc &  4.38 & -21.52 &   2.12 &   3.25 &   1.82 &   0.67 &   0.81 &   0.86 &   0.81 &   1.28 \\
 53 & 0.686 & 152.4 &  75.9 &  14 & Sc &  1.27 & -20.81 &   2.44 &   2.10 &   2.17 &   0.62 &   0.72 &    --- &   0.72 &   0.97 \\
 59 & 0.465 &  99.7 &  60.9 &   9 & S0a &  6.71 & -21.42 &   3.06 &   2.89 &   0.03 &   0.33 &   0.44 &   0.57 &   0.45 &   0.45 \\
 61 & 0.328 &  65.1 &  50.2 &  30 & Sa &  3.10 & -20.71 &   2.44 &   3.15 &   0.35 &   0.29 &   0.28 &    --- &   0.32 &   0.34 \\
 68 & 0.223 &  85.2 &  54.0 &  35 & E1 & 32.89 & -23.47 &   2.80 &   3.05 &   0.20 &   0.07 &   0.07 &    --- &   0.08 &   0.07 \\
 69 & 0.610 &  49.2 &  71.9 &  28 & Scd &  0.24 & -19.43 &   2.16 &   1.77 &   0.36 &   0.43 &   0.53 &    --- &   0.54 &   0.60 \\
 70 & 0.748 & 155.0 &  79.4 &  11 & Sb &  8.43 & -21.69 &   2.97 &   2.97 &   0.55 &   0.52 &   0.67 &   0.84 &   0.67 &   0.77 \\
 71 & 0.607 &  33.6 &  71.5 &  15 & Sc &  3.48 & -21.78 &   2.54 &   2.17 &   4.33 &   0.57 &   0.73 &   0.77 &   0.73 &   0.94 \\
 72 & 0.177 & 164.2 &  36.2 &  12 & S0 &  8.39 & -21.55 &   3.01 &   2.80 &   0.13 &   0.40 &   0.52 &   0.55 &   0.64 &   0.58 \\
 73 & 0.101 &  41.3 &  27.2 &  19 & Sb &  4.94 & -21.82 &   2.10 &   2.90 &   3.19 &   0.24 &   0.37 &    --- &   0.54 &   0.34 \\
 74 & 0.702 &  11.9 &  76.9 &   8 & Sa &  2.96 & -20.80 &   3.38 &   2.55 &   0.61 &   0.36 &   0.51 &   0.69 &   0.52 &   0.49 \\
 76 & 0.344 &  27.3 &  69.0 &  17 & E5 & 35.65 & -22.84 &   2.98 &   3.06 &   0.30 &   0.10 &   0.14 &    --- &   0.15 &   0.16 \\
 77 & 0.507 &  50.6 &  64.3 &  12 & Sa &  2.64 & -20.70 &   3.18 &   2.68 &   0.12 &   0.22 &   0.25 &   0.33 &   0.25 &   0.28 \\
 87 & 0.084 &  65.3 &  24.7 &  18 & S0a &  9.20 & -22.10 &   3.20 &   3.14 &   0.36 &   0.24 &   0.31 &    --- &   0.50 &   0.29 \\
100 & 0.318 & 153.8 &  49.4 &  14 & Sa &  1.05 & -19.94 &   3.17 &   2.85 &   1.09 &   0.22 &   0.22 &   0.24 &   0.25 &   0.28 \\
101 & 0.180 &  75.0 &  47.5 &  27 & E3 & 70.15 & -23.62 &   3.03 &   3.52 &   0.24 &   0.04 &   0.05 &    --- &   0.06 &   0.06 \\
103 & 0.575 &  96.8 &  69.0 &  10 & S0a &  5.53 & -21.42 &   3.10 &   3.00 &   0.10 &   0.42 &   0.53 &   0.68 &   0.54 &   0.54 \\
104 & 0.521 & 177.2 &  65.4 &  19 & S0a &  7.76 & -22.03 &   3.18 &   2.73 &   0.35 &   0.57 &   0.58 &    --- &   0.59 &   0.68 \\
108 & 0.390 &  99.1 &  55.2 &  13 & Sbc &  3.25 & -21.49 &   2.61 &   2.70 &   1.18 &   0.59 &   0.71 &   0.77 &   0.74 &   0.98 \\
115 & 0.587 & 153.5 &  69.8 &  19 & Sb &  3.19 & -20.80 &   2.04 &   2.84 &   3.10 &   0.56 &   0.70 &    --- &   0.71 &   1.01 \\
119 & 0.228 &  62.5 &  41.3 &  24 & S0 & 49.20 & -22.98 &   2.75 &   3.77 &   1.33 &   0.32 &   0.44 &    --- &   0.52 &   0.44 \\
127 & 0.119 &  50.6 &  29.6 &  17 & E6 &  8.47 & -22.44 &   2.59 &   3.28 &   0.65 &   0.25 &   0.38 &    --- &   0.54 &   0.37 \\
131 & 0.589 & 131.3 &  70.1 &  15 & Sab &  2.74 & -20.74 &   2.52 &   3.05 &   0.72 &   0.48 &   0.62 &    --- &   0.63 &   0.73 \\
134 & 0.568 &  84.4 &  68.7 &  13 & S0a &  6.78 & -21.81 &   3.23 &   2.77 &   0.35 &   0.44 &   0.54 &   0.75 &   0.55 &   0.57 \\
135 & 0.725 &  93.5 &  78.5 &  20 & Sa &  5.71 & -21.41 &   2.43 &  10.64 &   0.82 &   0.56 &   0.77 &    --- &   0.77 &   1.30 \\
143 & 0.710 & 137.0 &  77.7 &  12 & Scd &  0.38 & -19.46 &   2.51 &   1.69 &   0.63 &   0.44 &   0.56 &   0.60 &   0.57 &   0.65 \\
144 & 0.724 & 144.2 &  78.0 &  25 & Scd &  1.74 & -20.81 &   2.48 &   2.11 &   1.86 &   0.67 &   0.76 &    --- &   0.76 &   1.03 \\
146 & 0.475 &  87.7 &  61.7 &  15 & Sb &  2.69 & -21.16 &   2.26 &   2.54 &    --- &   0.52 &   0.70 &    --- &   0.71 &   0.89 \\
147 & 0.323 & 109.7 &  49.8 &  15 & Sbc &  3.34 & -21.55 &   2.14 &   2.54 &   2.81 &   0.46 &   0.65 &   0.75 &   0.70 &   0.80 \\
148 & 0.693 & 117.9 &  76.8 &  20 & Sc &  0.70 & -19.69 &   2.82 &   2.50 &   0.21 &   0.63 &   0.74 &    --- &   0.75 &   1.03 \\
149 & 0.360 &   9.1 &  52.8 &  18 & Sbc &  8.71 & -22.12 &   2.57 &   2.34 &   2.65 &   0.76 &   0.79 &   0.82 &   0.82 &   1.29 \\
150 & 0.698 &  44.0 &  77.1 &   9 & Sd &  0.17 & -19.23 &   2.89 &   1.36 &   0.50 &   0.33 &   0.34 &   0.35 &   0.34 &   0.37 \\
151 & 0.684 &  34.6 &  76.1 &  21 & Sb &  7.62 & -21.86 &   3.09 &   2.62 &   2.66 &   0.63 &   0.73 &    --- &   0.73 &   0.90 \\
152 & 0.569 & 120.9 &  68.6 &  17 & Sbc &  0.99 & -20.42 &   2.16 &   2.01 &   0.69 &   0.66 &   0.71 &    --- &   0.71 &   1.01 \\
153 & 0.781 & 134.3 &  80.7 &  15 & Sb &  4.78 & -21.19 &   2.36 &   2.47 &   2.24 &   0.60 &   0.79 &    --- &   0.79 &   1.18 \\
155 & 0.555 &  90.7 &  67.7 &  25 & Sb &  8.36 & -22.12 &   2.43 &   3.45 &   1.31 &   0.54 &   0.65 &    --- &   0.66 &   0.75 \\
156 & 0.519 & 135.7 &  65.4 &  16 & Sab &  7.28 & -21.68 &   2.74 &   2.97 &   0.94 &   0.62 &   0.76 &    --- &   0.77 &   1.03 \\
157 & 0.786 &  63.5 &  81.0 &  26 & Sc &  2.14 & -20.75 &   2.69 &   2.79 &   2.03 &   0.66 &   0.80 &    --- &   0.80 &   1.16 \\
159 & 0.769 & 167.9 &  80.4 &  24 & Sc &  1.41 & -20.43 &   2.32 &   2.56 &   1.15 &   0.73 &   0.85 &    --- &   0.85 &   1.49 \\
160 & 0.290 & 136.7 &  47.0 &   9 & E6 &  6.32 & -21.58 &   3.25 &   2.78 &   0.07 &   0.36 &   0.46 &   0.47 &   0.52 &   0.49 \\
161 & 0.432 & 167.2 &  58.6 &  35 & Sdm &  0.83 & -19.75 &   2.57 &   2.12 &   0.38 &   0.25 &   0.22 &    --- &   0.24 &   0.27 \\
162 & 0.159 &   6.5 &  34.9 &   9 & S0 &  4.83 & -20.80 &   3.33 &   2.75 &   0.05 &   0.34 &   0.49 &   0.64 &   0.62 &   0.50 \\
163 & 0.435 &  92.9 &  58.9 &  15 & Sab &  3.96 & -21.30 &   2.43 &   3.14 &    --- &   0.37 &   0.60 &    --- &   0.62 &   0.64 \\
164 & 0.146 & 132.5 &  32.6 &  28 & Sb &  5.90 & -22.19 &   1.99 &   3.32 &   1.30 &   0.44 &   0.61 &    --- &   0.73 &   0.71 \\
165 & 0.691 &  27.6 &  76.4 &  22 & Sbc &  5.82 & -21.81 &   2.31 &   2.55 &   8.55 &   0.77 &   0.84 &    --- &   0.85 &   1.43 \\
167 & 0.528 & 138.4 &  65.5 &   9 & Sa &  9.10 & -21.43 &   3.08 &   2.54 &   0.82 &   0.41 &   0.50 &   0.71 &   0.52 &   0.50 \\
171 & 0.269 & 174.5 &  60.4 &  26 & E2 & 34.59 & -22.86 &   3.08 &   3.14 &   0.10 &   0.08 &   0.10 &    --- &   0.11 &   0.10 \\
174 & 0.790 & 130.7 &  81.2 &  18 & Sab &  5.15 & -20.92 &    --- &   3.00 &   0.66 &   0.71 &   0.81 &   0.89 &   0.81 &   1.20 \\
183 & 0.394 & 131.5 &  55.6 &  14 & Sbc &  3.32 & -21.62 &   2.17 &   2.08 &   2.91 &   0.65 &   0.78 &   0.83 &   0.80 &   1.20 \\
185 & 0.682 &   3.7 &  75.9 &  11 & Sb &  1.37 & -20.29 &   2.58 &   2.43 &   0.48 &   0.53 &   0.69 &   0.83 &   0.70 &   0.86 \\
186 & 0.787 & 148.5 &  81.4 &  21 & Sab &  3.24 & -20.75 &    --- &   3.29 &   1.20 &   0.63 &   0.70 &    --- &   0.70 &   0.88 \\
187 & 0.141 & 113.9 &  32.8 &  24 & Sc &  1.84 & -21.29 &   2.06 &   3.96 &   1.11 &   0.50 &   0.68 &    --- &   0.80 &   0.94 \\
188 & 0.499 &  67.6 &  63.7 &   9 & Sb &  6.89 & -21.30 &   3.22 &   2.76 &   0.03 &   0.41 &   0.49 &    --- &   0.51 &   0.52 \\
189 & 0.192 & 160.4 &  37.9 &  19 & S0a & 16.33 & -22.59 &   2.97 &   2.99 &   0.50 &   0.29 &   0.36 &    --- &   0.44 &   0.36 \\
201 & 0.217 &  45.9 &  40.2 &   9 & E4 &  4.15 & -20.72 &   3.25 &   2.81 &   0.01 &   0.36 &   0.44 &   0.56 &   0.52 &   0.43 \\
209 & 0.118 &  48.4 &  29.4 &  26 & Sd &  0.46 & -20.24 &   1.98 &   1.95 &   0.57 &   0.21 &   0.29 &    --- &   0.42 &   0.34 \\
219 & 0.350 & 130.3 &  52.2 &  17 & Sa & 14.72 & -22.33 &   2.71 &   2.77 &   3.58 &   0.50 &   0.65 &   0.72 &   0.69 &   0.81 \\
231 & 0.793 &  31.7 &  81.2 &  32 & Sdm &  0.05 & -18.18 &   2.13 &   3.15 &   0.09 &   0.48 &   0.45 &    --- &   0.45 &   0.48 \\
232 & 0.115 &  80.8 &  29.2 &  24 & Scd &  1.31 & -20.94 &   1.88 &   1.95 &   1.46 &   0.48 &   0.56 &    --- &   0.72 &   0.72 \\
272 & 0.356 & 142.9 &  52.7 &  18 & E7 &  4.69 & -21.10 &   3.26 &   2.73 &   0.01 &   0.42 &   0.53 &    --- &   0.57 &   0.55 \\
273 & 0.791 & 162.9 &  81.2 &  25 & Sc &  2.48 & -21.05 &   2.51 &   3.22 &   1.43 &   0.79 &   0.84 &    --- &   0.85 &   1.42 \\
274 & 0.630 & 170.0 &  72.8 &  14 & Sab &  0.75 & -19.27 &   2.67 &   2.74 &   0.08 &   0.50 &   0.65 &    --- &   0.66 &   0.79 \\
275 & 0.437 &  82.6 &  59.1 &  19 & Sbc &  2.46 & -20.79 &   2.08 &   3.07 &   0.84 &   0.59 &   0.76 &    --- &   0.78 &   1.04 \\
277 & 0.356 &  19.9 &  52.3 &  26 & Sbc &  5.66 & -22.09 &   2.16 &   2.24 &   1.35 &   0.59 &   0.77 &    --- &   0.80 &   1.12 \\
278 & 0.595 & 138.1 &  70.3 &   9 & Sb &  7.74 & -22.12 &   2.81 &   2.48 &   4.87 &   0.32 &   0.47 &    --- &   0.47 &   0.54 \\
279 & 0.307 &  75.1 &  48.5 &  12 & E6 & 27.73 & -22.78 &   3.01 &   2.80 &   0.16 &   0.26 &   0.30 &   0.31 &   0.34 &   0.32 \\
281 & 0.738 &  41.9 &  78.7 &   8 & S0a & 12.62 & -21.76 &   3.44 &   2.93 &   0.12 &   0.44 &   0.52 &   0.71 &   0.53 &   0.53 \\
311 & 0.102 & 116.1 &  27.6 &  21 & Sab & 16.29 & -22.79 &   3.06 &   3.02 &   0.92 &   0.20 &   0.25 &    --- &   0.39 &   0.23 \\
312 & 0.269 &  23.2 &  44.9 &  32 & Sdm &  0.12 & -19.17 &   1.90 &   2.03 &   0.22 &   0.31 &   0.43 &    --- &   0.49 &   0.51 \\
314 & 0.785 &  61.4 &  81.2 &  13 & Sa &  6.38 & -21.36 &   3.10 &   2.69 &   1.11 &   0.58 &   0.68 &   0.81 &   0.68 &   0.81 \\
318 & 0.188 & 165.7 &  37.5 &  32 & E3 & 54.70 & -23.70 &   3.11 &   3.14 &   0.32 &   0.25 &   0.26 &    --- &   0.34 &   0.29 \\
319 & 0.756 & 140.8 &  79.6 &  15 & Sab &  8.99 & -21.51 &   2.71 &   2.79 &   0.79 &   0.49 &   0.65 &   0.81 &   0.65 &   0.70 \\
326 & 0.738 &  35.6 &  79.2 &  15 & Sb &  1.71 & -20.75 &   2.46 &   2.23 &   1.32 &   0.47 &   0.68 &    --- &   0.69 &   0.87 \\
339 & 0.459 & 173.3 &  60.5 &  13 & S0a &  5.78 & -21.41 &   2.69 &   2.92 &   0.02 &   0.41 &   0.56 &   0.76 &   0.59 &   0.62 \\
340 & 0.508 & 159.6 &  79.1 &  15 & S0a & 11.51 & -22.33 &   2.97 &   2.76 &   1.86 &   0.20 &   0.19 &    --- &   0.19 &   0.21 \\
341 & 0.329 &  60.4 &  50.6 &  12 & E6 & 20.75 & -22.35 &   3.35 &   2.76 &   0.86 &   0.39 &   0.48 &   0.54 &   0.52 &   0.49 \\
353 & 0.141 &  43.1 &  32.4 &  24 & Sd &  0.48 & -20.08 &   2.03 &   1.89 &   0.30 &   0.21 &   0.25 &    --- &   0.35 &   0.29 \\
361 & 0.742 &  15.1 &  78.9 &  12 & Sc &  0.16 & -18.32 &   2.78 &   1.90 &    --- &   0.16 &   0.17 &   0.34 &   0.17 &   0.20 \\
364 & 0.726 & 100.6 &  78.6 &  11 & Sa & 14.86 & -21.93 &   3.41 &   2.79 &   0.50 &   0.43 &   0.56 &   0.80 &   0.57 &   0.55 \\
381 & 0.635 & 117.0 &  73.1 &   9 & Sab &  8.09 & -22.07 &   2.91 &   2.60 &   0.86 &   0.54 &   0.61 &   0.77 &   0.62 &   0.70 \\
386 & 0.636 &  49.6 &  73.3 &  13 & Sab & 10.74 & -21.91 &   3.00 &   2.77 &   0.83 &   0.40 &   0.50 &   0.63 &   0.51 &   0.51 \\
387 & 0.416 &  42.9 &  57.6 &  15 & E5 & 24.15 & -22.98 &   3.34 &   2.94 &   0.12 &   0.31 &   0.35 &    --- &   0.37 &   0.35 \\
414 & 0.058 & 151.0 &  20.5 &  17 & Sb &  1.88 & -20.97 &   2.21 &   2.46 &   0.42 &   0.36 &   0.52 &    --- &   0.77 &   0.56 \\
436 & 0.233 & 171.5 &  41.7 &  21 & Sbc &  2.65 & -21.40 &   2.14 &   2.34 &   2.00 &   0.45 &   0.64 &    --- &   0.71 &   0.76 \\
437 & 0.499 &  67.8 &  63.8 &  14 & Sbc &  2.25 & -21.05 &   2.65 &   2.12 &    --- &   0.57 &   0.66 &   0.65 &   0.67 &   0.88 \\
476 & 0.492 &   8.1 &  63.5 &   9 & Sbc &  2.65 & -21.22 &   2.72 &   1.89 &   3.71 &   0.54 &   0.67 &   0.71 &   0.69 &   0.84 \\
479 & 0.279 & 168.8 &  46.1 &  14 & S0a & 11.83 & -22.00 &   2.77 &   2.75 &   1.32 &   0.49 &   0.60 &    --- &   0.66 &   0.72 \\
486 & 0.507 &  12.1 &  64.4 &  14 & Scd &  0.28 & -19.91 &   2.72 &   1.48 &   0.69 &   0.62 &   0.71 &   0.63 &   0.73 &   1.00 \\
489 & 0.258 &  97.6 &  44.0 &  16 & Sbc &  4.86 & -21.90 &   2.29 &   2.17 &   3.42 &   0.61 &   0.66 &    --- &   0.73 &   0.94 \\
500 & 0.632 & 151.3 &  72.9 &  16 & Sbc &  2.26 & -21.14 &   2.31 &   2.43 &   1.23 &   0.71 &   0.83 &    --- &   0.83 &   1.40 \\
502 & 0.597 &  85.4 &  70.7 &  18 & Sa &  2.30 & -20.63 &   2.62 &   2.30 &   0.42 &   0.56 &   0.72 &    --- &   0.73 &   0.93 \\
515 & 0.334 & 164.4 &  50.7 &  30 & Sbc &  4.69 & -21.88 &   1.96 &   3.16 &   0.89 &   0.75 &   0.80 &    --- &   0.83 &   1.34 \\
518 & 0.244 &  97.7 &  42.7 &  21 & Sb &  1.93 & -20.98 &   1.87 &   2.55 &   0.55 &   0.64 &   0.79 &    --- &   0.84 &   1.25 \\
548 & 0.325 & 179.5 &  49.9 &  15 & Sc &  0.98 & -20.72 &   2.01 &   1.56 &   0.87 &   0.37 &   0.50 &    --- &   0.55 &   0.59 \\
569 & 0.706 &  57.0 &  77.3 &  12 & Sb &  2.82 & -21.03 &   2.78 &   2.40 &   0.80 &   0.58 &   0.73 &   0.83 &   0.74 &   0.96 \\
577 & 0.847 &   2.3 &  83.5 &  38 & Sdm &  6.50 & -22.17 &   2.80 &   2.41 &   2.09 &   0.49 &   0.54 &    --- &   0.54 &   0.62 \\
580 & 0.475 &  41.4 &  62.0 &  17 & Sbc &  2.05 & -21.05 &   2.04 &   2.38 &    --- &   0.73 &   0.76 &    --- &   0.78 &   1.19 \\
588 & 0.309 &  80.6 &  65.4 &  30 & E1 & 32.06 & -23.03 &   2.87 &   2.91 &   0.18 &   0.07 &   0.08 &    --- &   0.08 &   0.08 \\
589 & 0.107 &  42.9 &  28.0 &  20 & E3 & 35.16 & -22.83 &   3.22 &   2.84 &   0.21 &   0.08 &   0.10 &    --- &   0.17 &   0.12 \\
592 & 0.235 &  46.5 &  55.7 &  55 & E0 & 49.54 & -24.11 &   2.72 &   3.00 &   0.13 &   0.05 &   0.06 &    --- &   0.07 &   0.07 \\
593 & 0.681 &  53.2 &  75.8 &  16 & Sa & 11.64 & -22.56 &   2.49 &   3.17 &   6.40 &   0.34 &   0.44 &    --- &   0.45 &   0.44 \\
602 & 0.166 &  39.6 &  35.0 &   9 & E1 &  9.68 & -22.38 &   2.97 &   2.53 &   0.11 &   0.30 &   0.38 &   0.46 &   0.48 &   0.40 \\
603 & 0.313 & 100.6 &  48.8 &  15 & Scd &  0.48 & -20.25 &   2.46 &   1.59 &   0.64 &   0.45 &   0.51 &   0.48 &   0.56 &   0.61 \\
606 & 0.664 &  73.6 &  74.9 &  19 & Sd &  0.14 & -19.28 &   2.25 &   1.58 &   0.39 &   0.37 &   0.40 &    --- &   0.41 &   0.40 \\
607 & 0.454 & 133.4 &  60.5 &   8 & S0 & 13.65 & -22.76 &   3.32 &   2.39 &   0.14 &   0.40 &   0.51 &   0.59 &   0.53 &   0.54 \\
608 & 0.241 &   1.3 &  42.9 &  16 & Sbc &  5.37 & -21.81 &   2.19 &   2.42 &   2.04 &   0.28 &   0.36 &    --- &   0.43 &   0.37 \\
611 & 0.233 &  57.4 &  41.7 &  17 & Sbc &  1.72 & -21.06 &   2.28 &   1.90 &   0.90 &   0.51 &   0.65 &    --- &   0.73 &   0.89 \\
612 & 0.397 & 149.2 &  73.2 &  25 & E6 & 31.77 & -23.28 &   3.02 &   2.88 &   0.17 &   0.07 &   0.09 &    --- &   0.09 &   0.10 \\
614 & 0.436 &   3.4 &  58.7 &  15 & Sc &  3.02 & -21.82 &   2.46 &   2.10 &    --- &   0.71 &   0.80 &   0.83 &   0.82 &   1.30 \\
630 & 0.354 & 169.0 &  52.2 &  19 & Sbc &  0.73 & -20.12 &   2.42 &   2.43 &   0.10 &   0.51 &   0.66 &   0.72 &   0.70 &   0.85 \\
633 & 0.325 &  33.2 &  50.0 &  20 & E0 &  3.20 & -21.07 &   2.95 &   2.71 &   0.08 &   0.19 &   0.20 &    --- &   0.23 &   0.22 \\
634 & 0.135 & 101.4 &  33.1 &  18 & Sab &  4.49 & -21.43 &   2.48 &   2.85 &   2.30 &   0.35 &   0.45 &    --- &   0.58 &   0.52 \\
657 & 0.437 &  21.5 &  59.2 &  30 & Sdm &  0.26 & -19.75 &   2.14 &   1.67 &   0.19 &   0.45 &   0.46 &    --- &   0.48 &   0.53 \\
663 & 0.649 & 105.7 &  74.3 &  19 & Sab & 18.20 & -22.37 &   2.76 &   3.07 &   2.08 &   0.57 &   0.70 &    --- &   0.71 &   0.83 \\
664 & 0.704 & 116.6 &  77.0 &  15 & Sb &  1.76 & -20.30 &   2.41 &   2.41 &   0.22 &   0.52 &   0.71 &    --- &   0.72 &   0.89 \\
665 & 0.403 & 160.7 &  56.3 &  15 & Sb & 11.38 & -22.37 &   2.47 &   2.77 &   1.51 &   0.41 &   0.62 &    --- &   0.65 &   0.67 \\
676 & 0.216 &  86.7 &  40.3 &  24 & Sb &  3.83 & -21.27 &   2.42 &   2.88 &   0.10 &   0.42 &   0.58 &    --- &   0.66 &   0.60 \\
684 & 0.293 & 111.4 &  47.2 &  20 & Sb & 18.75 & -22.57 &   2.31 &   2.83 &   1.72 &   0.52 &   0.72 &    --- &   0.76 &   0.88 \\
707 & 0.182 &  41.9 &  37.3 &  25 & Scd &  1.38 & -20.76 &   1.93 &   1.87 &   1.21 &   0.48 &   0.61 &    --- &   0.72 &   0.78 \\
708 & 0.321 & 174.5 &  49.7 &  31 & E5 & 10.57 & -21.95 &   2.79 &   2.82 &   0.02 &   0.15 &   0.19 &    --- &   0.21 &   0.20 \\
714 & 0.537 &  12.6 &  66.4 &  14 & Sbc &  3.82 & -21.85 &   2.39 &   1.83 &   4.47 &   0.73 &   0.82 &    --- &   0.83 &   1.41 \\
715 & 0.486 &  63.1 &  62.9 &  12 & Sbc &  0.73 & -20.18 &   3.03 &   2.28 &   0.82 &   0.53 &   0.64 &   0.79 &   0.66 &   0.80 \\
740 & 0.190 & 128.2 &  37.5 &  18 & Sa & 19.86 & -22.77 &   3.05 &   2.83 &   1.89 &   0.17 &   0.19 &    --- &   0.25 &   0.20 \\
744 & 0.058 &  30.7 &  20.6 &  19 & S0 &  8.47 & -21.74 &   3.21 &   2.72 &   0.07 &   0.14 &   0.11 &    --- &   0.22 &   0.15 \\
748 & 0.256 &  16.9 &  43.9 &  13 & Sbc &  1.82 & -20.94 &   2.15 &   1.84 &   1.39 &   0.51 &   0.64 &   0.68 &   0.71 &   0.87 \\
749 & 0.705 &  93.7 &  77.1 &  22 & Sdm &  0.47 & -20.37 &   2.51 &   1.51 &   0.93 &   0.50 &   0.62 &    --- &   0.62 &   0.75 \\
754 & 0.630 & 164.0 &  72.4 &  10 & Sbc &  1.92 & -20.98 &   3.06 &   2.35 &   1.65 &   0.56 &   0.63 &   0.73 &   0.64 &   0.75 \\
758 & 0.764 & 125.6 &  80.3 &  26 & Scd &  0.25 & -19.39 &    --- &   1.75 &   0.26 &   0.62 &   0.68 &    --- &   0.68 &   0.83 \\
764 & 0.435 & 131.0 &  58.7 &  16 & Sbc &  7.05 & -22.29 &   2.36 &   2.44 &   1.78 &   0.52 &   0.71 &    --- &   0.74 &   0.86 \\
768 & 0.476 &  43.2 &  62.1 &  14 & Sbc &  0.85 & -20.46 &   2.30 &   1.94 &   0.86 &   0.69 &   0.73 &   0.74 &   0.74 &   1.01 \\
769 & 0.337 & 132.0 &  51.0 &  21 & Sbc &  1.61 & -20.95 &   1.84 &   2.13 &   0.97 &   0.61 &   0.72 &    --- &   0.76 &   1.06 \\
774 & 0.790 & 140.7 &  81.2 &  20 & Sb & 16.60 & -22.64 &   3.33 &   3.17 &   2.22 &   0.60 &   0.71 &    --- &   0.71 &   0.82 \\
775 & 0.703 &  34.3 &  77.1 &  21 & Sc &  2.07 & -20.96 &   2.29 &   2.38 &   1.67 &   0.73 &   0.78 &    --- &   0.78 &   1.18 \\
778 & 0.134 &  19.2 &  31.4 &  13 & S0 & 16.44 & -22.61 &   3.36 &   2.90 &   1.28 &   0.25 &   0.33 &   0.40 &   0.46 &   0.36 \\
780 & 0.453 & 130.1 &  60.1 &  18 & E7 &  7.01 & -22.13 &   3.14 &   2.54 &   0.04 &   0.43 &   0.51 &    --- &   0.53 &   0.53 \\
781 & 0.511 &  82.0 &  78.6 &  37 & E4 & 21.23 & -23.21 &   2.95 &   2.91 &   0.15 &   0.08 &   0.08 &    --- &   0.08 &   0.09 \\
783 & 0.725 & 138.2 &  78.3 &  18 & Sb &  0.99 & -19.99 &   2.37 &   2.47 &   0.80 &   0.63 &   0.76 &    --- &   0.76 &   1.10 \\
787 & 0.592 &  51.4 &  70.3 &  12 & S0a &  7.96 & -21.41 &   3.68 &   2.79 &   0.11 &   0.29 &   0.35 &   0.59 &   0.36 &   0.33 \\
789 & 0.382 & 150.7 &  54.8 &  16 & Sb & 16.07 & -22.74 &   2.07 &   2.87 &   2.30 &   0.57 &   0.76 &    --- &   0.79 &   1.03 \\
791 & 0.359 & 154.0 &  52.8 &  34 & Sa & 16.71 & -22.17 &   2.69 &   3.80 &   2.63 &   0.60 &   0.71 &    --- &   0.75 &   0.99 \\
795 & 0.537 & 161.8 &  66.5 &  16 & Sab &  4.30 & -21.36 &   2.77 &   2.70 &   2.23 &   0.59 &   0.72 &   0.78 &   0.73 &   0.92 \\
796 & 0.591 &  24.1 &  70.4 &  13 & Sb &  8.75 & -21.80 &   2.29 &   2.65 &   1.24 &   0.52 &   0.66 &   0.76 &   0.66 &   0.81 \\
797 & 0.746 & 126.0 &  79.2 &  21 & Sb &  1.25 & -20.38 &   2.44 &   2.51 &   1.19 &   0.56 &   0.71 &    --- &   0.71 &   0.90 \\
798 & 0.695 & 101.0 &  76.7 &  16 & Sbc &  1.98 & -20.71 &   2.50 &   2.29 &   0.89 &   0.64 &   0.80 &    --- &   0.80 &   1.22 \\
801 & 0.102 &  43.3 &  27.4 &  10 & Sa &  3.01 & -21.09 &   3.12 &   2.15 &   2.59 &   0.31 &   0.24 &   0.23 &   0.38 &   0.32 \\
804 & 0.555 & 132.4 &  68.0 &  12 & Sb &  2.07 & -20.80 &   2.96 &   2.28 &   0.12 &   0.38 &   0.51 &   0.62 &   0.52 &   0.53 \\
806 & 0.386 &  83.9 &  54.8 &  18 & E4 & 10.21 & -22.08 &   2.83 &   2.83 &   0.10 &   0.21 &   0.25 &    --- &   0.27 &   0.25 \\
807 & 0.411 & 161.0 &  57.1 &  15 & Sb &  8.99 & -21.96 &   2.24 &   2.95 &   0.70 &   0.49 &   0.62 &    --- &   0.65 &   0.71 \\
809 & 0.655 &  62.7 &  74.7 &  33 & Sa & 16.22 & -22.15 &   3.01 &   3.08 &   0.10 &   0.44 &   0.57 &    --- &   0.57 &   0.55 \\
810 & 0.596 &  11.7 &  70.2 &  17 & Sbc &  5.25 & -21.81 &   2.32 &   2.38 &   5.71 &   0.70 &   0.79 &   0.85 &   0.80 &   1.21 \\
813 & 0.205 &  91.3 &  39.4 &  22 & Sbc &  4.86 & -21.86 &   1.97 &   3.55 &   1.41 &   0.50 &   0.66 &    --- &   0.75 &   0.82 \\
814 & 0.178 & 114.1 &  36.4 &  14 & E5 & 28.71 & -22.73 &   3.29 &   2.86 &   0.59 &   0.26 &   0.31 &    --- &   0.40 &   0.30 \\
815 & 0.306 & 133.5 &  65.1 &  19 & E4 & 10.02 & -22.08 &   3.03 &   3.14 &   0.04 &   0.04 &   0.06 &    --- &   0.06 &   0.06 \\
816 & 0.273 & 157.1 &  45.4 &   9 & E5 & 10.14 & -21.88 &   3.28 &   2.92 &   0.10 &   0.27 &   0.30 &   0.34 &   0.35 &   0.32 \\
817 & 0.626 &  28.9 &  72.5 &  22 & Scd &  0.70 & -20.26 &   2.14 &   2.01 &   0.55 &   0.61 &   0.68 &    --- &   0.69 &   0.91 \\
818 & 0.775 &  53.6 &  80.5 &  18 & Sab &  3.32 & -20.55 &   2.49 &   2.79 &   1.65 &   0.60 &   0.73 &    --- &   0.73 &   0.99 \\
820 & 0.382 &   0.1 &  55.1 &  27 & Sbc &  3.37 & -21.30 &   1.96 &   3.21 &   0.56 &   0.59 &   0.68 &    --- &   0.71 &   0.92 \\
821 & 0.566 & 102.0 &  68.9 &  28 & Sb &  8.59 & -22.23 &   2.22 &   2.93 &   3.26 &   0.67 &   0.78 &    --- &   0.79 &   1.11 \\
822 & 0.380 & 133.6 &  54.1 &  19 & S0a &  9.93 & -22.32 &   2.55 &   2.79 &   1.76 &   0.29 &   0.42 &    --- &   0.45 &   0.46 \\
823 & 0.443 & 156.6 &  59.4 &  20 & Sbc &  1.38 & -20.55 &   1.89 &   2.47 &   0.47 &   0.75 &   0.83 &    --- &   0.85 &   1.60 \\
824 & 0.570 & 157.3 &  68.9 &  20 & Sb &  4.67 & -21.46 &   2.12 &   2.35 &   1.46 &   0.64 &   0.82 &    --- &   0.83 &   1.43 \\
825 & 0.776 & 162.2 &  80.5 &  20 & Sbc &  1.21 & -20.47 &   2.54 &   2.10 &    --- &   0.75 &   0.82 &    --- &   0.82 &   1.31 \\
826 & 0.587 & 128.2 &  69.7 &  12 & S0a & 13.12 & -21.95 &   3.39 &   2.96 &   0.60 &   0.32 &   0.40 &   0.57 &   0.41 &   0.39 \\
827 & 0.826 & 179.9 &  82.8 &  18 & Sc &  0.29 & -19.11 &   2.36 &   2.02 &   0.19 &   0.46 &   0.60 &   0.75 &   0.60 &   0.70 \\
828 & 0.761 & 140.8 &  79.8 &  19 & Sc &  0.77 & -20.43 &   2.41 &   1.76 &   2.23 &   0.39 &   0.48 &   0.56 &   0.49 &   0.51 \\
829 & 0.043 &   4.8 &  17.7 &  21 & E1 & 24.21 & -22.86 &   3.19 &   2.77 &   0.06 &   0.09 &   0.10 &    --- &   0.23 &   0.10 \\
830 & 0.716 &  63.9 &  77.8 &  17 & Sb & 10.79 & -22.17 &   2.40 &   2.68 &   3.04 &   0.73 &   0.82 &    --- &   0.82 &   1.41 \\
831 & 0.644 & 125.0 &  73.7 &  14 & Sbc &  1.63 & -21.04 &   2.48 &   2.07 &   1.71 &   0.68 &   0.73 &   0.79 &   0.74 &   1.06 \\
832 & 0.242 &  73.3 &  42.6 &  15 & E5 & 42.56 & -23.48 &   3.36 &   2.75 &   1.12 &   0.23 &   0.26 &    --- &   0.31 &   0.25 \\
834 & 0.791 & 108.2 &  81.3 &  12 & Sb & 10.21 & -21.85 &   2.78 &   2.86 &   1.64 &   0.40 &   0.59 &   0.66 &   0.59 &   0.68 \\
835 & 0.447 &  58.7 &  59.9 &  11 & E7 & 26.67 & -22.65 &   2.91 &   2.88 &   0.09 &   0.37 &   0.51 &   0.65 &   0.53 &   0.51 \\
837 & 0.729 &  92.8 &  78.7 &  11 & Sb &  1.87 & -20.73 &   2.56 &   1.97 &   4.59 &   0.67 &   0.80 &   0.86 &   0.80 &   1.21 \\
838 & 0.696 & 128.4 &  76.8 &  11 & Sa &  6.56 & -21.19 &   3.11 &   2.67 &   0.12 &   0.45 &   0.58 &   0.78 &   0.58 &   0.60 \\
840 & 0.371 & 144.5 &  71.4 &  38 & E6 & 53.09 & -23.85 &   3.15 &   2.92 &   0.21 &   0.10 &   0.09 &    --- &   0.10 &   0.12 \\
841 & 0.769 & 110.2 &  80.3 &  26 & Sc &  0.73 & -20.00 &   2.02 &   2.24 &   0.71 &   0.44 &   0.67 &    --- &   0.67 &   0.81 \\
842 & 0.228 &  49.6 &  41.3 &  20 & Sb &  3.71 & -21.24 &   2.44 &   2.77 &   2.09 &   0.47 &   0.60 &    --- &   0.68 &   0.66 \\
843 & 0.782 &  22.3 &  80.5 &  23 & Scd &  0.20 & -19.32 &    --- &   2.46 &   0.50 &   0.25 &   0.33 &    --- &   0.33 &   0.22 \\
844 & 0.416 & 126.4 &  57.3 &  11 & S0a &  8.30 & -21.49 &   3.40 &   3.11 &   0.06 &   0.28 &   0.41 &   0.63 &   0.43 &   0.37 \\
845 & 0.371 & 103.8 &  71.2 &  22 & E7 & 32.14 & -23.39 &   3.28 &   2.99 &   0.16 &   0.09 &   0.13 &    --- &   0.14 &   0.14 \\
846 & 0.354 & 110.8 &  69.8 &  24 & E5 & 19.95 & -22.70 &   2.80 &   2.78 &   0.55 &   0.06 &   0.06 &    --- &   0.07 &   0.07 \\
847 & 0.652 & 147.2 &  74.4 &  20 & Sb &  9.68 & -22.12 &   2.83 &   2.73 &    --- &   0.65 &   0.75 &    --- &   0.75 &   1.00 \\
848 & 0.720 &  69.9 &  78.1 &  23 & Sb &  3.64 & -20.99 &   2.33 &   2.78 &   0.44 &   0.58 &   0.78 &    --- &   0.79 &   1.08 \\
849 & 0.395 & 108.5 &  55.8 &  24 & Sbc & 10.42 & -22.76 &   2.04 &   2.22 &   3.51 &   0.82 &   0.88 &    --- &   0.89 &   1.96 \\
850 & 0.466 & 173.4 &  61.1 &  12 & Sab & 16.26 & -22.46 &   3.34 &   2.66 &   0.59 &   0.29 &   0.35 &   0.42 &   0.36 &   0.32 \\
851 & 0.375 &  15.1 &  71.5 &  28 & E5 & 49.09 & -23.48 &   2.81 &   3.27 &   0.47 &   0.09 &   0.09 &    --- &   0.09 &   0.10 \\
852 & 0.416 &  59.0 &  57.6 &  20 & Scd &  0.28 & -19.56 &   2.41 &   1.80 &   0.26 &   0.19 &   0.24 &    --- &   0.26 &   0.28 \\
854 & 0.662 &  91.6 &  75.2 &  12 & Sb &  7.48 & -21.92 &   2.69 &   2.48 &   1.22 &   0.55 &   0.66 &   0.82 &   0.67 &   0.79 \\
856 & 0.358 &  89.1 &  52.7 &  17 & Sb &  2.00 & -21.18 &   2.19 &   2.32 &   1.49 &   0.69 &   0.76 &    --- &   0.79 &   1.18 \\
857 & 0.640 &  42.6 &  73.6 &  14 & Sbc &  7.87 & -21.87 &   2.10 &   2.60 &   3.51 &   0.70 &   0.81 &    --- &   0.81 &   1.29 \\
858 & 0.555 & 173.1 &  67.8 &  15 & S0a & 40.46 & -22.92 &   3.47 &   2.95 &   0.64 &   0.50 &   0.60 &    --- &   0.61 &   0.64 \\
859 & 0.355 &  65.3 &  70.0 &  34 & E4 & 12.08 & -22.42 &   3.00 &   3.23 &   0.06 &   0.09 &   0.08 &    --- &   0.09 &   0.09 \\
860 & 0.569 &  34.6 &  68.7 &   8 & S0 &  5.64 & -21.37 &   3.25 &   2.64 &   0.05 &   0.34 &   0.39 &   0.55 &   0.40 &   0.37 \\
861 & 0.780 &  54.1 &  80.6 &  24 & Sbc &  2.66 & -21.22 &   2.34 &   2.57 &   0.83 &   0.71 &   0.85 &    --- &   0.85 &   1.40 \\
862 & 0.632 &  34.2 &  73.2 &  23 & Sc & 10.33 & -22.57 &   2.05 &   2.40 &   7.00 &   0.67 &   0.82 &    --- &   0.82 &   1.25 \\
863 & 0.513 & 112.3 &  64.6 &  13 & Sab & 10.89 & -22.09 &   2.66 &   3.16 &   1.45 &   0.43 &   0.57 &   0.79 &   0.59 &   0.62 \\
864 & 0.349 &  12.0 &  69.2 &  19 & E3 & 15.60 & -22.73 &   3.13 &   2.85 &   0.14 &   0.04 &   0.06 &    --- &   0.06 &   0.07 \\
865 & 0.326 & 178.5 &  50.6 &  12 & S0 & 12.39 & -22.10 &   3.42 &   2.87 &   0.76 &   0.26 &   0.38 &   0.57 &   0.42 &   0.36 \\
867 & 0.716 & 119.0 &  77.9 &   9 & Sab &  2.42 & -20.46 &   2.94 &   2.65 &   0.10 &   0.45 &   0.59 &   0.72 &   0.59 &   0.63 \\
868 & 0.638 & 159.5 &  73.4 &  17 & Sb &  4.79 & -21.69 &   2.18 &   2.56 &   2.64 &   0.77 &   0.85 &    --- &   0.86 &   1.48 \\
869 & 0.256 & 131.1 &  43.9 &  20 & Sb &  8.77 & -22.39 &   2.12 &   2.91 &   2.13 &   0.54 &   0.74 &    --- &   0.80 &   0.95 \\
870 & 0.360 & 127.1 &  53.0 &  13 & S0 & 24.49 & -21.91 &   2.96 &   2.89 &   0.00 &   0.34 &   0.51 &   0.68 &   0.55 &   0.54 \\
871 & 0.575 & 126.8 &  69.1 &  18 & Sb & 10.79 & -22.15 &   2.41 &   2.85 &   1.21 &   0.73 &   0.83 &    --- &   0.83 &   1.31 \\
872 & 0.217 &  63.5 &  40.4 &  19 & Sab &  3.70 & -21.38 &   2.12 &   3.16 &   0.29 &   0.45 &   0.62 &    --- &   0.70 &   0.71 \\
873 & 0.456 &  57.3 &  60.2 &  16 & Sb & 12.39 & -22.56 &   2.51 &   3.76 &   2.04 &   0.49 &   0.57 &    --- &   0.59 &   0.68 \\
874 & 0.325 &  39.9 &  50.1 &  13 & S0a & 33.65 & -22.73 &   3.27 &   3.21 &   0.51 &   0.32 &   0.38 &   0.51 &   0.42 &   0.39 \\
876 & 0.450 & 107.7 &  60.1 &  18 & Sbc &  6.18 & -21.83 &   2.04 &   3.07 &   1.82 &   0.71 &   0.83 &    --- &   0.85 &   1.49 \\
877 & 0.610 &  37.1 &  71.6 &  17 & Sab &  6.95 & -21.84 &    --- &   3.31 &   2.50 &   0.50 &   0.60 &   0.65 &   0.60 &   0.68 \\
878 & 0.779 &  32.1 &  80.6 &  24 & Scd &  0.11 & -18.77 &   2.29 &   1.72 &   0.35 &   0.47 &   0.57 &    --- &   0.57 &   0.69 \\
881 & 0.300 &  18.5 &  47.8 &  17 & E3 & 27.86 & -23.05 &   3.27 &   2.93 &   0.15 &   0.35 &   0.38 &    --- &   0.43 &   0.39 \\
885 & 0.739 &  47.4 &  79.1 &  15 & Sc &  0.10 & -18.41 &   2.67 &   1.67 &   0.17 &   0.18 &   0.41 &   0.51 &   0.41 &   0.44 \\
886 & 0.473 &   9.9 &  61.7 &  12 & Sa & 11.72 & -22.45 &   3.06 &   2.90 &   2.76 &   0.51 &   0.64 &   0.76 &   0.66 &   0.72 \\
887 & 0.324 &  14.4 &  49.8 &  15 & Sbc &  8.53 & -22.48 &   2.39 &   2.59 &   5.20 &   0.60 &   0.76 &   0.82 &   0.80 &   1.12 \\
888 & 0.107 &   0.3 &  35.4 &  36 & E1 & 23.39 & -23.47 &   3.05 &   3.50 &   0.23 &   0.06 &   0.06 &    --- &   0.09 &   0.06 \\
889 & 0.320 &  65.0 &  49.5 &  12 & Sab &  7.62 & -22.04 &   3.11 &   2.51 &   4.37 &   0.51 &   0.61 &   0.62 &   0.65 &   0.75 \\
890 & 0.561 & 159.2 &  68.6 &  12 & Sb &  4.68 & -21.56 &   2.83 &   2.75 &   0.81 &   0.46 &   0.66 &   0.78 &   0.67 &   0.78 \\
892 & 0.772 &  70.8 &  80.4 &  23 & Sb &  2.32 & -20.96 &   3.11 &   3.23 &   2.05 &   0.68 &   0.75 &    --- &   0.75 &   0.98 \\
893 & 0.153 &  41.9 &  43.4 &  27 & E2 & 82.04 & -23.50 &   2.86 &   3.42 &   0.26 &   0.11 &   0.12 &    --- &   0.15 &   0.12 \\
894 & 0.678 & 143.1 &  76.0 &  17 & Sa & 14.19 & -22.08 &   2.72 &   3.44 &   0.32 &   0.51 &   0.65 &    --- &   0.65 &   0.72 \\
895 & 0.794 & 118.0 &  81.4 &  27 & Scd &  0.11 & -18.88 &   2.45 &   2.01 &   0.20 &   0.43 &   0.55 &    --- &   0.55 &   0.60 \\
896 & 0.623 &  25.7 &  72.8 &  13 & Sbc &  5.60 & -21.86 &   2.69 &   2.53 &   4.35 &   0.55 &   0.64 &   0.77 &   0.64 &   0.74 \\
898 & 0.472 & 160.6 &  61.8 &  20 & Sbc &  3.17 & -22.07 &   2.28 &   2.41 &   7.00 &   0.76 &   0.78 &    --- &   0.80 &   1.33 \\
900 & 0.095 & 154.0 &  33.3 &  24 & E4 & 27.04 & -22.89 &   2.89 &   3.13 &   0.33 &   0.05 &   0.08 &    --- &   0.13 &   0.08 \\
901 & 0.603 &  16.1 &  71.1 &  20 & Sbc &  3.97 & -21.75 &   2.13 &   2.81 &   2.60 &   0.55 &   0.63 &    --- &   0.63 &   0.78 \\
902 & 0.468 & 149.6 &  61.4 &   9 & Sa &  9.18 & -21.54 &   3.52 &   3.13 &   0.06 &   0.23 &   0.31 &   0.46 &   0.32 &   0.29 \\
903 & 0.324 &  83.0 &  67.0 &  20 & E4 & 17.66 & -22.54 &   3.07 &   3.11 &   0.07 &   0.05 &   0.10 &    --- &   0.11 &   0.09 \\
904 & 0.455 & 150.2 &  60.4 &  16 & Sbc &  5.75 & -21.91 &   2.97 &   2.63 &   7.37 &   0.61 &   0.66 &    --- &   0.68 &   0.82 \\
905 & 0.666 &  37.5 &  75.3 &  20 & Sd &  0.28 & -19.67 &   2.38 &   2.02 &    --- &   0.49 &   0.49 &    --- &   0.49 &   0.59 \\
906 & 0.545 &  25.1 &  67.1 &  17 & Sc &  1.89 & -20.61 &    --- &   2.71 &   1.68 &   0.57 &   0.68 &    --- &   0.69 &   0.86 \\
907 & 0.732 &  18.5 &  78.5 &  20 & Sbc &  1.24 & -20.00 &   2.26 &   2.77 &   0.49 &   0.67 &   0.82 &    --- &   0.82 &   1.31 \\
908 & 0.550 & 134.9 &  67.2 &  11 & S0 &  7.93 & -21.32 &   3.52 &   2.83 &   0.01 &   0.38 &   0.54 &    --- &   0.56 &   0.60 \\
909 & 0.701 & 157.9 &  76.9 &  21 & Sc &  1.09 & -20.56 &   2.30 &   2.67 &   1.96 &   0.68 &   0.77 &    --- &   0.78 &   1.31 \\
910 & 0.644 &  23.9 &  73.8 &  17 & Sb &  1.80 & -19.45 &   2.79 &   3.18 &   0.06 &   0.67 &   0.91 &    --- &   0.91 &   4.78 \\
911 & 0.174 &  50.2 &  46.7 &  35 & E3 &  8.79 & -22.69 &   3.25 &   3.10 &   0.10 &   0.14 &   0.13 &    --- &   0.16 &   0.15 \\
912 & 0.300 &   7.9 &  47.9 &  10 & S0 &  9.57 & -21.47 &   3.41 &   2.83 &   0.03 &   0.19 &   0.27 &   0.44 &   0.31 &   0.25 \\
913 & 0.044 &  10.9 &  17.8 &  14 & Sa &  1.33 & -20.26 &   2.61 &   2.43 &   1.59 &   0.43 &   0.49 &   0.56 &   0.80 &   0.59 \\
914 & 0.615 &  76.0 &  72.1 &  17 & Sb &  3.38 & -21.10 &   2.56 &   2.72 &   0.88 &   0.63 &   0.77 &    --- &   0.78 &   1.12 \\
915 & 0.182 & -11.0 &  36.7 &  12 & Sb &  3.16 & -21.58 &   2.73 &   2.28 &   2.68 &   0.49 &   0.57 &   0.63 &   0.68 &   0.66 \\
916 & 0.372 & 133.9 &  53.9 &  11 & S0 &  9.04 & -21.76 &   3.22 &   2.90 &   0.11 &   0.34 &   0.47 &   0.59 &   0.50 &   0.46 \\
917 & 0.477 & 138.5 &  62.2 &  14 & S0 & 10.45 & -21.74 &   3.16 &   3.07 &   0.09 &   0.42 &   0.49 &   0.64 &   0.51 &   0.51 \\
919 & 0.656 &  22.2 &  74.5 &   9 & S0 &  9.68 & -21.69 &   3.06 &   2.90 &   0.43 &   0.39 &   0.58 &   0.75 &   0.58 &   0.58 \\
920 & 0.211 & 171.9 &  39.9 &  28 & Sbc &  1.64 & -21.34 &   1.97 &   2.92 &    --- &   0.65 &   0.66 &    --- &   0.75 &   0.98 \\
923 & 0.546 &  92.1 &  67.3 &  15 & E7 & 11.30 & -22.02 &   3.29 &   2.86 &   0.11 &   0.35 &   0.43 &   0.50 &   0.44 &   0.41 \\
924 & 0.194 &  31.6 &  38.0 &  21 & Sb &  2.45 & -21.04 &   2.73 &   2.57 &   0.49 &   0.48 &   0.57 &    --- &   0.67 &   0.64 \\
925 & 0.273 & 148.8 &  45.6 &  21 & Sab & 17.58 & -22.05 &   2.63 &   3.25 &   0.84 &   0.40 &   0.43 &    --- &   0.50 &   0.48 \\
926 & 0.820 &  75.5 &  82.4 &  17 & Sc &  0.76 & -19.77 &   2.07 &   1.93 &   0.64 &   0.60 &   0.79 &    --- &   0.79 &   1.39 \\
927 & 0.587 &  34.7 &  70.2 &  14 & Sb & 12.00 & -22.23 &   2.96 &   2.81 &  15.20 &   0.39 &   0.42 &   0.61 &   0.43 &   0.49 \\
929 & 0.610 &  56.0 &  71.5 &  20 & Sbc &  5.43 & -22.01 &   2.00 &   2.61 &   3.79 &   0.69 &   0.84 &    --- &   0.84 &   1.40 \\
930 & 0.502 & 140.4 &  63.6 &  16 & Sc &  0.66 & -20.63 &   2.37 &   1.97 &   1.05 &   0.57 &   0.66 &    --- &   0.68 &   0.85 \\
932 & 0.535 & 120.2 &  66.3 &  15 & Sa & 28.51 & -22.59 &   3.04 &   2.92 &   0.48 &   0.48 &   0.55 &    --- &   0.56 &   0.60 \\
933 & 0.715 & 104.3 &  77.6 &  11 & Sab &  4.18 & -21.17 &   2.56 &   2.71 &   2.46 &   0.50 &   0.58 &    --- &   0.59 &   0.69 \\
934 & 0.722 &  35.4 &  77.8 &  19 & Sbc &  0.56 & -19.49 &   2.47 &   2.37 &   0.40 &   0.43 &   0.48 &    --- &   0.49 &   0.53 \\
935 & 0.611 & 110.0 &  71.7 &  27 & Sc &  1.13 & -20.69 &   2.39 &   2.78 &   0.97 &   0.34 &   0.50 &    --- &   0.51 &   0.51 \\
937 & 0.607 &  44.8 &  71.1 &  32 & Ir &  0.19 & -19.56 &   2.44 &   1.51 &   0.40 &   0.18 &   0.19 &    --- &   0.19 &   0.22 \\
2999 & 0.154 &  61.6 &  33.8 &  13 & Sbc &  3.48 & -21.56 &   2.36 &   2.65 &   1.86 &   0.44 &   0.62 &   0.74 &   0.74 &   0.70 \\
4034 & 0.438 &  43.4 &  59.1 &  15 & S0 &  7.18 & -22.09 &   2.01 &   2.83 &   1.90 &   0.26 &   0.34 &    --- &   0.36 &   0.36 \\
\enddata
\tablecomments{Col.~1: Galaxy name. Col.~2: average ellipticity measured in the outer parts of the galaxy, using SDSS images. Col.~3: average position angle measured in the outer parts of the galaxy, using SDSS images. Col.~4: statistical inclination (see Appendix~\ref{app:deproj}). Col.~5: effective radii (in arcsec) of the galaxy, measured as described in \citet{Walcher_etal_2014}. Col.~6: Hubble type of the galaxy from \citet{Walcher_etal_2014}. Col.~7: total stellar mass of the galaxy, measured as described in \citet{Walcher_etal_2014}. Col.~8: total absolute magnitude in $r-$band from SDSS \citep{DR7}. Col.~9: concentration index (ratio of Petrosian radius rad90 and rad50). Col.~10: SDSS Petrosian $u-r$ color. Col.~11: star formation rate based on extinction corrected H$\alpha$ measurements \citep{sanchez17}. Col.~12,13,14: \lamR\ measured on an elliptical aperture with semi-major axis 0.5\,\Reff, \Reff, and 2\,\Reff\ respectively. Col.~15: deprojected \lamRe\ ($\lamReintr$, see Appendix~\ref{app:deproj}). Col.~16: \VSe\ measured on an elliptical aperture with semi-major axis \Reff. We refer the reader to FLV17 for further properties of the galaxies not listed here.}
\label{tab:sample}
\end{deluxetable}

%============================================================================================

% END DOCUMENT
%=====================================================================
% \label{LastPage}
\end{document}